\documentclass[11pt]{article}
\pdfoutput=1
\usepackage[utf8]{inputenc}
\usepackage{jheppub}
\usepackage{amsmath,amssymb,amsfonts,graphicx,slashed,color,amsthm,mathtools,upgreek, enumerate, tensor, scalerel, subfig}
\usepackage{arydshln}
\usepackage[dvipsnames]{xcolor}
\usepackage{bbold}
\allowdisplaybreaks

\title{Canonical Purification and the Quantum Extremal Shock}

\author[a]{\!Onkar Parrikar,} 
\author[a]{\!Vivek Singh}
\affiliation[\,a]{Department of Theoretical Physics,
Tata Institute of Fundamental Research,\\ Mumbai 400005, India.}
\newcommand{\ch}{\chi}
\newcommand{\tch}{\tilde{\chi}}
\newcommand{\st}{\star}
\newcommand{\beq}{\begin{equation}}
\newcommand{\eeq}{\end{equation}}
\newcommand{\beqn}{\begin{eqnarray}}
\newcommand{\eeqn}{\end{eqnarray}}
\newcommand{\pa}{\partial}
\newcommand{\bs}{\boldsymbol}

\newcommand{\cM}{\mathcal{M}}
\newcommand{\cN}{\mathcal{N}}

\newcommand{\cH}{\mathcal{H}}
\newcommand{\cO}{\mathcal{O}}
\newcommand{\cU}{\mathcal{R}}
\newcommand{\cA}{\mathcal{A}}
\newcommand{\cl}{\lambda}

\newcommand{\Th}{\Theta}

\begin{document}

\abstract{We study the canonical purification of pure, bi-partite states (with respect to one of the parties) obtained by turning on sources in the Euclidean path integral. In holographic conformal field theories, the Lorentzian bulk dual of the canonical purification consists of the corresponding entanglement wedge glued to its CPT image at the quantum extremal surface. However, the mismatch in the classical expansions at the QES due to quantum corrections needs to be supported by a shock in the bulk matter stress tensor in order for the bulk to satisfy Einstein's equations. Working perturbatively to first order in double-trace sources around the thermofield double state, we demonstrate that the state of the bulk matter in the dual to the canonically purified boundary CFT state precisely has this quantum extremal shock in the bulk stress tensor. We interpret our results as the emergence of gravitational physics from the CFT entanglement structure in a context where bulk quantum corrections are important.   

}

\maketitle

\parskip=12pt

\section{Introduction}
Consider a general, full-rank, bi-partite state $\Psi$ in the (for the moment, finite dimensional) tensor product Hilbert space $\cH_{L}\otimes \cH_R$. Any such state can always be written in the form:
\beq
|\Psi\rangle = \sum_{n} \sqrt{p_n}\, |\tch_n\rangle_L\otimes |\ch_n\rangle_R,
\eeq
where $p_n$ are the eigenvalues of the reduced density matrices on $L$ and $R$, $\ch_n$ form an orthonormal set of eigenstates of the reduced density matrix on the right, and $\tch_n$ form an orthonormal set of eigenstates of the reduced density matrix on the left. While the eigenvalues are common to both the parties, the eigenstates are not. If we only had access to, say, the left factor, then we could write down a purification for the density matrix as follows:
\beq
|\Psi^{\st}\rangle  = \sum_{n} \sqrt{p_n}\, |\tch_n\rangle_L\otimes |\tch_n^{\st}\rangle_{L^{\st}}.
\eeq
Here 
\beq
|\tch_n^{\st}\rangle = \Theta |\tch_n\rangle,
\eeq
where $\Theta$ is an anti-unitary operator on $L$.\footnote{In quantum field theory, we could take it to be CPT in even dimensions or CRT in odd dimensions, with R being reflection along one spatial direction. \cite{Witten:2018zxz}} This new state is called the \emph{canonical purification} of $\Psi$ with respect to the left side \cite{Dutta:2019gen}.\footnote{Equivalently, the canonical purification of a density matrix $\rho$ is defined as the state $|\sqrt{\rho}\rangle$ viewed as a vector in the Hilbert space $\text{End}(\cH) = \cH\otimes \cH^*$.} Note that $\Psi^{\st}$ resembles the thermofield double state. Physically, if one only had access to the left party in $\Psi$ and not to the right, then we can think of $\Psi^{\st}$ as the ``simplest'' purification that one could build from this information. Since $\Psi$ and $\Psi^{\st}$ are two different purifications of the same reduced density matrix on $L$, these two states are related by a unitary transformation on the right:\footnote{Here, we are using the full-rank condition. More generally, the two purifications would be related by an isometry.}
\beq
|\Psi^{\st}\rangle := \cU_{\Psi}|\Psi\rangle,
\eeq
where 
\beq
\cU_{\Psi}: \cH_R \to \cH_{L^{\st}},\;\;\;\cU_{\Psi} := \sum_n  |\tch^{\st}_n\rangle_{L^{\st}}\langle \ch_n|_R.
\eeq
The operator $\cU_{\Psi}$ quantifies how ``complex'' it is to reconstruct the original state $\Psi$ from $\Psi^{\st}$. We should emphasize that entanglement or R\'enyi entropies between the left and the right parties are blind to $\cU_{\Psi}$. It therefore seems interesting to study aspects of this operator $\cU_{\Psi}$, which goes beyond entanglement in quantifying properties of the state $\Psi$. We will call this operator $\cU_{\Psi}$ which maps $\Psi$ to its canonical purification with respect to $L$ the \emph{reflection operator} with respect to $L$. When $\Psi$ is full-rank, the reflection operator is uniquely specified by the condition that it maps $\Psi$ to its canonical purification.

There are several motivations to study the reflection operator in holographic conformal field theories: one important motivation comes from the quantum error correction perspective on the bulk to boundary map in AdS/CFT \cite{Papadodimas:2013jku, Almheiri:2014lwa, Dong:2016eik, Harlow:2016vwg}. It was shown by Harlow \cite{Harlow:2016vwg} that for a bulk degree of freedom (say, a qudit) within the entanglement wedge of a boundary subregion $A$, the encoding map $V$ into the dual CFT takes the general form:
\beq
|\psi_i\rangle_{\text{CFT}} = V|i\rangle_{\text{bulk}} = U_{A}|i\rangle_{A_1}\otimes |\chi\rangle_{A_2,\bar{A}},
\eeq
where $\{|i\rangle\}$ forms a basis of states for the bulk qudit, and $\cH_A= \cH_{A_1}\otimes \cH_{A_2} \oplus \cH_{A_3}$, with $\cH_{A_1}$ being the same dimension as that of the code subspace. Harlow's structure theorem is a general consequence of the Ryu-Takayanagi formula \cite{Ryu:2006bv, Hubeny:2007xt, Faulkner:2013ana}. Importantly, we can think of the unitary $U_A$ appearing in Harlow's structure theorem as a reflection operator: introduce an auxiliary reference system $``\text{ref}"$ which has the same dimension as that of the code subspace, and consider the maximally entangled state:
\beq
|\Psi\rangle = \frac{1}{\sqrt{d_{\text{code}}}}\sum_i |i\rangle_{\text{ref}}\otimes |\psi_i\rangle_{\text{CFT}}.
\eeq
Then, the unitary $U_A$ is precisely the adjoint of the reflection operator for this state $\Psi$ with respect to $\text{ref}\cup \bar{A}$. Furthermore, this particular reflection operator gives a simple recipe for bulk reconstruction: we can represent any bulk operator $\phi$ on the code subspace as a boundary operator on $A$ via the formula
\beq
O_A = U_{A}\phi_{A_1} U_{A}^{\dagger}.
\eeq
Thus, the reflection operator finds a natural role in formulating the bulk-to-boundary map in AdS/CFT as a quantum error correcting code. 

A second (perhaps much more direct) motivation, comes from the fact that the reflection operator is closely linked with the canonical purification, which finds several interesting applications in holography. For instance, the canonical purification is crucially used in identifying the area of the outermost extremal surface as the simple entropy \cite{Engelhardt:2018kcs, Engelhardt:2021mue}.  Along similar lines, the reflected entropy \cite{Dutta:2019gen} for a mixed two-party state $\rho_{AB}$ is also defined in terms of the canonical purification $\Psi^{\st}_{ABA^{\st}B^{\st}}$ as the entanglement entropy of $AA^{\st}$. The reflected entropy is an interesting information theoretic quantity \cite{Akers:2019gcv, Chandrasekaran:2020qtn, Zou:2020bly, Hayden:2021gno, Akers:2021pvd, Akers:2022max}, one which finds a natural bulk dual in terms of the cross section area of the entanglement wedge of $AB$. Finally, it was recently argued in \cite{Engelhardt:2022qts} that for a black hole evaporating into a non-gravitational bath, the canonical purification of the total state with respect to the black hole side is dual to a connected wormhole, thus realizing the ER=EPR idea in the context of an evaporating black hole (see also \cite{Geng:2020fxl, Anderson:2020vwi} for other approaches). While the original state of the radiation plus the evaporating black hole does not appear to have a wormhole in it, the state after the action of the corresponding reflection operator does; in this way, the  reflection operator in this case acts to ``geometrize" the entanglement in the originally complex and non-geometric state. 

\begin{figure}
    \centering
\begin{tabular}{c c c}
\includegraphics[height=4cm]{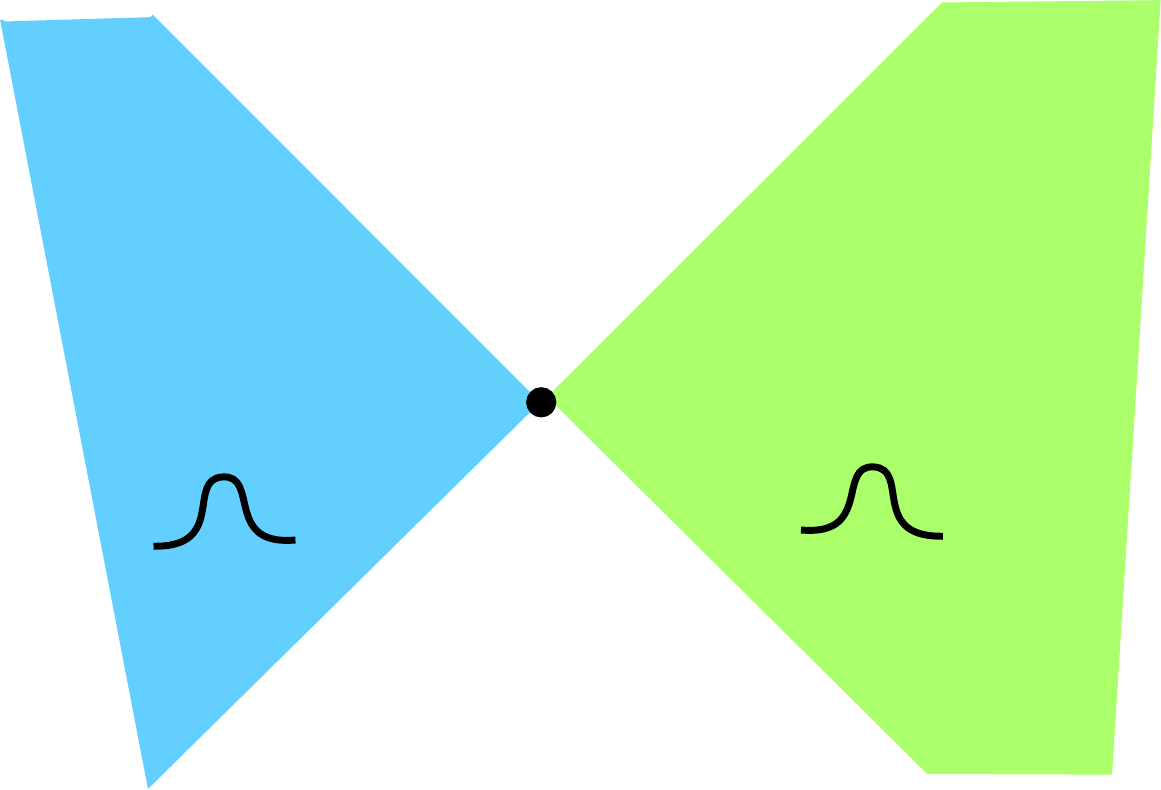} & \hspace{1cm} &  \includegraphics[height=4cm]{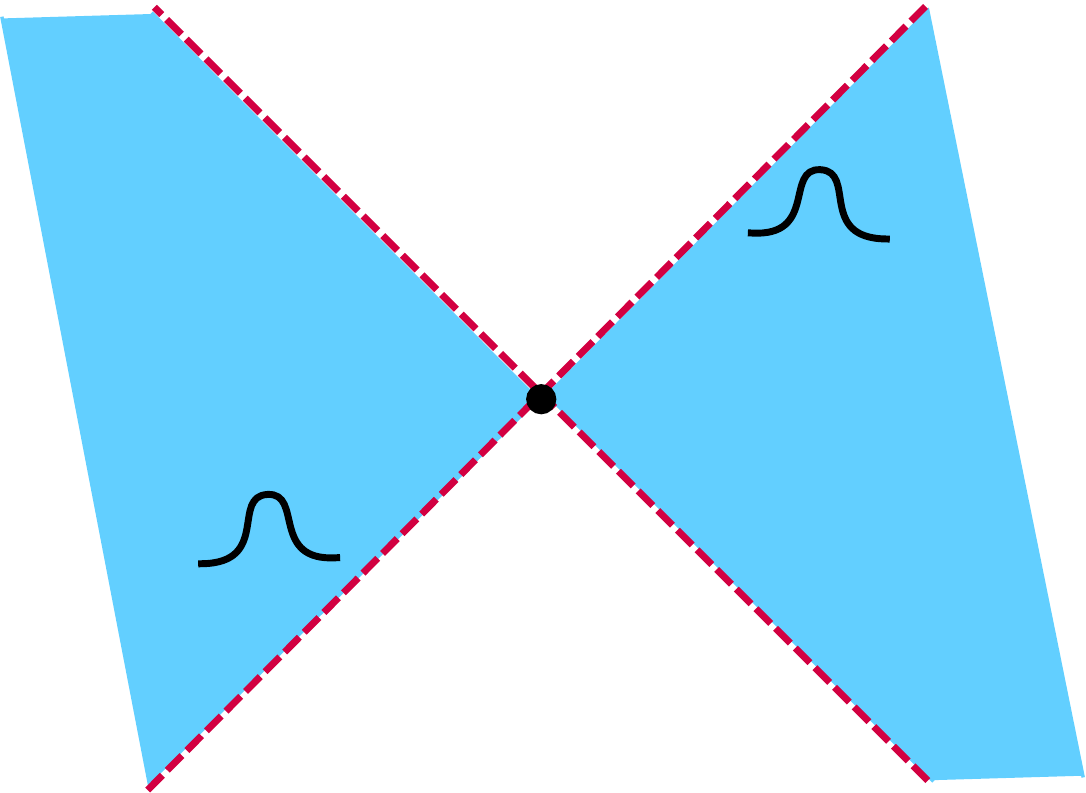}
\end{tabular}
\caption{(Left) A portion of the bulk geometry dual to some holographic state $\Psi$. The entanglement wedge of the left party is shown in blue and the entanglement wedge of the right party is shown in green. (Right) The Engelhardt-Wall proposal for the geometry dual to the canonical purification $\Psi^{\st}$ with respect to the left party consists of the left entanglement wedge glued to its CPT image at the quantum extremal surface. In situations where the quantum extremal surface is not classically extremal, the geometry needs to be supported by a shock (red dashed lines) in the bulk matter stress tensor. }
    \label{fig:EW}
\end{figure}

For holographic theories, it was proposed by Engelhardt and Wall \cite{Engelhardt:2018kcs} that the classical, Lorentzian bulk geometry dual to the canonical purification with respect to a boundary subregion $A$ is obtained by taking the entanglement wedge of $A$ and gluing it to its CPT image at the RT or HRRT surface \cite{Ryu:2006bv, Hubeny:2007xt} dual to $A$ (see figure \ref{fig:EW}). A replica trick argument for this proposal was later given by Dutta and Faulkner \cite{Dutta:2019gen} (see also \cite{Marolf:2019zoo}). In gluing together portions of solutions of Einstein equations to obtain new solutions, one must impose junction conditions at the gluing surface in order to ensure that the resulting geometry also satisfies Einstein's equations. In the case at hand, the fact that the co-dimension two surface we are gluing across is a classically extremal surface implies that these junction conditions are trivially satisfied (see section \ref{sec:EWconstruction} for more details). The resulting geometry contains an entire Cauchy surface, and one can obtain the full solution by evolving the data on this surface with the Einstein equations. Upon including quantum corrections, the gluing must be done across the quantum extremal surface (QES) \cite{Engelhardt:2014gca}. However, due to quantum corrections, the QES is not generically classically extremal, and now the junction conditions imply that the bulk matter must be in a state whose stress tensor has a delta function ``shock''\cite{Bousso:2019dxk} proportional to the first shape-derivative of the bulk entanglement entropy, in order for Einstein's equations to be satisfied. 

Our goal in this paper will be to study the reflection operator in a perturbative setup. The main application we have in mind is to verify the above prediction of general relativity for the bulk stress tensor shock in the context of the Engelhardt-Wall construction. We will consider a family of states $\Psi_{\cl}$ labelled by some parameter $\cl$. We will first derive a differential equation for $\cU_{\cl} \equiv \cU_{\Psi_{\cl}}$ along the flow parametrized by $\cl$; this equation will involve more familiar quantities such as the modular Hamiltonian and modular flow. In order to be concrete, we will then apply this general equation to the thermofield double (TFD) state perturbed by turning on a source (with a small amplitude) in the Euclidean path integral. In a holographic quantum system, we will then use this to compute the bulk stress tensor one-point function (to first order in the deformation) in the bulk dual to the canonically purified state and show that it has the quantum extremal shock contribution required for the Engelhardt-Wall construction to work. While we will explicitly demonstrate the existence of this shock to first order in perturbation theory around the TFD state, we expect that with some mild assumptions, our calculation can be extended beyond perturbation theory (i.e., at finite deformation parameter $\cl$). Since the shock is a prediction of Einstein's equations from the bulk point of view, we are seeing here the emergence of bulk gravitational physics from the CFT entanglement structure \cite{Lashkari:2013koa, Faulkner:2013ica, Jacobson:2015hqa, Faulkner:2017tkh, Haehl:2017sot, Lewkowycz:2018sgn}, but in a context where quantum corrections in the bulk are important (see also \cite{Haehl:2019fjz, Belin:2021htw} for related previous work). 

The rest of the paper is organized as follows: in section \ref{sec:EWconstruction}, we review the Engelhardt-Wall construction of the bulk dual to the canonical purification. In section \ref{sec:flow}, we study the reflection operator for a general one-parameter family of states. We then apply this to the special case of the TFD state deformed by a source in the Euclidean path integral, and derive an explicit formula for the reflection operator in this context to first order in perturbation theory. In section \ref{sec:gravity}, we apply this formula to holographic quantum systems in order to study the one-point function of the bulk matter stress tensor and demonstrate the existence of the quantum extremal shock. We end in section \ref{sec:discussion} with some concluding remarks and open directions. 

\section{Review of Engelhardt-Wall construction}\label{sec:EWconstruction}
In this section, we briefly review the construction of Engelhardt and Wall (EW) for the holographic dual of the canonical purification of a bi-partite state. The EW geometry is a Lorentzian geometry constructed in the following way: let us begin with the original Lorentzian spacetime $M$ dual to the original state $\Psi$. Let $\sigma$ be the quantum extremal surface (QES) corresponding to the left subregion, and let $D_{\sigma}$ be the corresponding entanglement wedge. The EW proposal for the geometry dual to the canonical purification with respect to the left is to glue $D_{\sigma}$ to its CPT image at the surface $\sigma$, then evolve the resulting data on a Cauchy slice using Einstein's equations to obtain the full Lorentzian geometry. However, for this to work, we must impose a set of co-dimension two junction conditions on the geometric data at $\sigma$ in $M$. These junction conditions are analogous to, and in fact follow from, the standard, co-dimension one junction conditions which are imposed when gluing two solutions to Einstein's equations across a co-dimension one hypersurface \cite{Israel:1966rt, Barrabes:1991ng}. 

The basic idea is as follows: let us imagine, for the moment, that we have two different spacetimes $M$ and $M'$ with some Cauchy slices $\Sigma$ and $\Sigma'$ respectively. Now, consider co-dimension two surfaces $\sigma$ and $\sigma'$ in $M$ and $M'$ respectively, which divide $\Sigma$ and $\Sigma'$ into two parts. Let us call one part $\text{In}_{\Sigma}(\sigma)$ and the other $\text{Out}_{\Sigma}(\sigma)$ in $M$, we can write similar divisions of the Cauchy slice in $M'$. This procedure naturally divides each spacetime into four parts, namely, $I_W(\sigma) \equiv D[\text{In}_{\Sigma}(\sigma)]$, $O_W(\sigma) \equiv D[\text{Out}_{\Sigma}(\sigma)]$, $J^+[\sigma]$ and $J^-[\sigma]$, where $D$ denotes the domain of dependence and $J^{+(-)}$ denotes the causal future (past). We have a similar division for $M'$ as well. We wish to glue $I_W(\sigma)$ to $O_W(\sigma')$ by identifying the two surfaces $\sigma$ and $\sigma'$. For this to work, the most basic thing we must demand is that the intrinsic geometry on $\sigma$ and $\sigma'$ should be identical, or more precisely, the induced metrics $h= h_{ij}dy^idy^j$ on the two surfaces should be equivalent (up to a change of coordinates) -- this is the first junction condition. 

Next, let us imagine that there exists a consistent solution to Einstein's equations which contains $V_{\text{in}} \equiv I_W(\sigma)$ and $V_{\text{out}} \equiv O_W(\sigma')$ glued together at $\sigma=\sigma'$. Let us consider the null surface $\cN_k$ which separates $V_{\text{out}}\cup J^-(\sigma)$ from $V_{\text{in}} \cup J^+(\sigma)$. Let $k$ be the generating vector field tangent to null geodesics (not necessarily affinely parametrized) along $\cN_k$; at $\sigma$, we can take $k$ to be orthogonal to $\sigma$. Let $\ell^{\mu}$ be a transverse null vector field satisfying $\ell.k=-1$ everywhere on $\mathcal{N}_k$. We can take $\ell$ such that at $\sigma$ it is orthogonal to $\sigma$ and agrees with the generating vector field of the null surface $\cN_{\ell}$ separating $V_{\text{out}} \cup J^+(\sigma)$ and $V_{\text{in}} \cup J^-(\sigma)$. The idea is to now apply the co-dimension one Barrab\`es-Israel junction conditions \cite{Barrabes:1991ng, Israel:1966rt} individually to $\cN_k$ and $\cN_{\ell}$. For instance, the junction condition across $\cN_k$ gives the following expression for the matter stress tensor localized to this null sheet:
\beq
8\pi G_N \;T_{\mu\nu}^{(k)}= -\left(\left[\theta_{(\ell)}\right] k_{\mu}k_{\nu} + [{\chi_{(\ell)}}_{(\mu}]k_{\nu)}+ [\kappa_{(\ell)}] h_{\mu\nu}\right)\delta(\cN_k),
\eeq
where $\theta_{(\ell)}$ is the expansion of the null-geodesic congruence generated by the vector field $\ell$, ${\chi_{(\ell)}}_{\mu}$ is called its twist, and $\kappa_{(\ell)}$ measures the in-affinity of the geodesic congruence generated by $k$:
\beqn
\theta_{(\ell)} &=& h^{ij} h^{\mu}_i h^{\nu}_j \nabla_{\mu} \ell_{\nu},\\
\chi_{(\ell)\,\mu}&=&\frac{1}{2} h^{\mu}_i k^{\nu} \nabla_{\mu}\ell_{\nu},\\
\kappa_{(\ell)}&=&-k^{\nu} k^{\mu} \nabla_{\mu} \ell_{\nu}=\ell_{\nu}k^{\mu}\nabla_{\mu}k^{\nu}.
\eeqn
Finally, the notation $[\cdot]$ stands for difference across $\cN_k$. We can write a similar equation for $\cN_{\ell}$ as well. We are interested in evaluating these constraints at $\sigma$. Since the in-affinity at a point along a geodesic (in the present case, corresponding to where it intersects $\sigma$) can be adjusted by an arbitrary rescaling, we can set the discontinuity in the in-affinity to zero at $\sigma$ by a suitable choice of parametrization. Furthermore, for our specific case where we wish to glue an entanglement wedge to its CPT image, the twist term also drops out, since the twist is even under CPT. On the other hand, the expansion is odd under CPT, and so we get
\beq\label{IJC}
8\pi G_N T^{(k)}_{\mu\nu} = -2\theta_{(\ell)} k_{\mu}k_{\nu}\,\delta(\cN_k)\;\;\;\;\;\;\; (\text{at}\;\sigma).
\eeq
For classically extremal surfaces, the expansion vanishes and the gluing does not require any singular matter stress tensor. However, for a quantum extremal surface, the expansion is not zero, but given by the quantum extremality formula \cite{Engelhardt:2014gca}: 
\beq
\theta_{(\ell)} = -\frac{4G_N}{\sqrt{h}} \ell^{\mu}\frac{\delta S_{\text{bulk}}}{\delta x^{\mu}}.
\eeq
Thus, general relativity makes a prediction for the singular part of the matter stress tensor at the quantum extremal surface in the Lorentzian geometry dual to the canonical purification of a holographic state:
\beq 
2\pi T^{(k)}_{\mu\nu} = \frac{2}{\sqrt{h}} \ell^{\mu}\frac{\delta S_{\text{bulk}}}{\delta x^{\mu}}\,k_{\mu}k_{\nu}\delta(\mathcal{N}_k),
\eeq 
with a similar prediction for the stress tensor localized to $\mathcal{N}_{\ell}$. In principle, we need to compute the state of bulk matter fields in the bulk dual to the canonically purified state, evaluate the corresponding bulk stress tensor, and check whether it satisfies the above prediction. Our goal is to do this in the perturbative framework.

\section{Perturbation theory for the reflection operator}\label{sec:flow}
Consider a bi-partite Hilbert space $\cH_L\otimes \cH_R$, where $\cH_L$ and $\cH_R$ are both finite dimensional Hilbert spaces of the same dimension. Let us say that we have a general one-parameter family of states $\Psi_{\cl} \in \cH_L\otimes \cH_R$ which are all full rank. At any value of $\cl$, we can construct the reduced density matrices $\rho_L(\cl)$ and $\rho_R(\cl)$ corresponding to the left and right factors respectively. Accordingly, we have the one-parameter family of modular Hamiltonians $K_L(\cl)$ and $K_R(\lambda)$, where the modular Hamiltonian for a density matrix $\rho$ is defined as $K=-\log\,\rho$. At any given value of $\cl$, we have a Schmidt decomposition for the state $\Psi_{\cl}$:
\beq
|\Psi_{\cl}\rangle = \sum_n e^{-\frac{1}{2}E_n(\cl)}|\tch_n(\cl)\rangle_L\otimes |\ch_n(\cl)\rangle_R.
\eeq
In terms of the modular Hamiltonians, the $\ch_n$ and $\tch_n$ satisfy
\beq
K_R(\cl)|\ch_n(\cl)\rangle_R = E_n(\cl) |\ch_n(\cl)\rangle_R,
\eeq
\beq
K_L(\cl)|\tch_n(\cl)\rangle_L = E_n(\cl) |\tch_n(\cl)\rangle_L,
\eeq
where note that the eigenvalues are common to both sides. In terms of these quantities, recall that the reflection operator $\cU_{\cl}$ is defined as:
\beq
\cU_{\cl} = \sum_n |\tch^{\st}\rangle_{L^{\st}}\langle \ch_n|_R,
\eeq
where $|\tch^{\st}\rangle_{L^{\st}} = \Theta |\tch\rangle_{L^{\st}}$, and $\Theta$ is an anti-unitary operator which we will take to be CPT. Our first goal is to derive a differential equation for $\cU_{\cl}$ along the flow parametrized by $\cl$. 
\subsection{Flow equation}
Upon an infinitesimal deformation of the parameter $\lambda$, the change in the eigenstates of, say $K_R$, is given by
\beq
\frac{d}{d\lambda} |\ch_n\rangle_R = \sum_{m\neq n} \frac{\langle \ch_m | \frac{d}{d\lambda} K_R |\ch_n\rangle_R}{(E_n(\lambda)-E_m(\lambda))} |\ch_m\rangle_R.
\eeq
Here we have assumed that the eigenvalues are non-degenerate. We can rewrite this in the following way:
\beqn
\frac{d}{d\lambda} |\ch_n\rangle_R &=& \sum_{m\neq n} \int_0^{\infty}idt\,e^{-\epsilon t}\left(\langle \ch_m | e^{it K_R(\lambda)} \frac{d}{d\lambda} K_R  e^{-it K_R(\cl)}|\ch_n\rangle_R\right)|\ch_m\rangle_R\nonumber\\
&=& \int_0^{\infty}idt\,e^{-\epsilon t} e^{it K_R(\cl)} \frac{d}{d\lambda} K_R  e^{-it K_R(\cl)}|\ch_n\rangle_R-\frac{i}{\epsilon}\frac{d}{d\lambda} E_n(\cl)\,|\ch_n\rangle_R.
\eeqn
Here we have introduced a regulator $\epsilon \to 0^+$, which plays two roles: firstly, it regulates the $t$-integral at large $t$ in the first line. Secondly, it allows us to add and subtract the $m=n$ term in the sum, which together with 
$$\sum_m |\ch_m\rangle\langle \ch_m|_R = \mathbb{1}_R,$$ 
allows us to rewrite the expression as in the second line. Note that the $\frac{1}{\epsilon}$ divergence is not really present, since it cancels the corresponding divergence from the first term; we have merely chosen to write the expression in this way for convenience.  A similar formula is also true for the modular eigenstates of the left party, and so we get the following flow equations for the eigenstates:
\beq \label{transport}
\frac{d}{d\lambda}|\chi_n\rangle_R = i \cA_{R}|\chi_n\rangle_R,\;\;\; \frac{d}{d\lambda}|\tch_n\rangle_L = i \cA_{L}|\tch_n\rangle_L,
\eeq
where
\beq \label{conn0R}
\cA_R(\cl) = a_R(\cl)+\int_0^{\infty}dte^{-\epsilon t}\,e^{itK^{(\cl)}_R}\frac{d}{d\lambda} K^{(\cl)}_R  e^{-it K^{(\lambda)}_R},
\eeq
\beq \label{conn0L}
\cA_L(\cl) = a_L(\cl)+\int_0^{\infty}dte^{-\epsilon t}\,e^{itK^{(\cl)}_L}\frac{d}{d\lambda} K^{(\cl)}_L  e^{-it K^{(\lambda)}_L}.
\eeq
Here $a_L$ and $a_R$ are the diagonal terms proportional to $\frac{1}{\epsilon}$. There is an important subtlety we need to address at this point: orthonormality does not fix the overall phase of an eigenstate of the modular Hamiltonian, i.e., we have the freedom $\ch_n \to e^{i\phi_n}\ch_n$. So, as far as the eigenstates of the modular Hamiltonian are concerned, the diagonal terms in the above flow equation are ambigious. Some of this ambiguity is fixed by the fact that we want $\ch_n(\cl)$ and $\tch_n(\cl)$ to be a Schmidt basis for the family of states $\Psi(\cl)$. In particular, the sum of the left and the right phases $(\phi_n + \tilde{\phi}_n)$ is fixed, but the relative phase $(\phi_n - \tilde{\phi}_n)$ is not; this is good enough for our purposes, because the reflection operator is unambiguous once the Schmidt condition is imposed. Crucially, these ambiguities all correspond to diagonal terms in the modular eigenstate basis, and for what we are interested in, we will not need to worry about fixing them. We will simply gather all these diagonal terms inside $a_L$ and $a_R$ henceforth.\footnote{More precisely, $(a_L+a_R)$ can be fixed by the Schmidt condition. But as we will see later, $a_L$ and $a_R$ will drop out of the calculations we are interested in.} 

Coming back to the reflection operator, the change in $\cU_{\cl}$ in these terms is given by
\beq\label{diff}
i\frac{d}{d\lambda} \cU_{\lambda} = i\sum_n \Big(\Th\frac{d}{d\lambda} |\tch_n\rangle_{L^{\st}}\langle \ch_n|_R + \Th|\tch_n\rangle_{L^{\st}}\frac{d}{d\lambda} \langle \ch_n|_R\Big)=\cA^{\st}_{L}(\cl)\,\cU_{\cl}+\cU_{\cl}\,\cA_{R}(\cl),
\eeq
%Happily, the $\frac{1}{\epsilon}$ terms cancel between the two sides. Thus, we are left with a simple differential equation for the operator $\cU_{\cl}$:
%\beq \label{diff}
%-i\frac{d}{d\cl}\cU_{\cl} = \cA_{L^{\st}}(\cl)\,\cU_{\cl}-\cU_{\cl}\,\cA_{R}(\cl),
%\eeq
where we have defined 
\beq
\cA^{\st}_{L}(\cl) = \Th\,\cA_{L}(\cl)\,\Th^{-1}.
\eeq
While we have focused on the special case with only one parameter $\lambda$, the formulas above apply naturally to the more general case where the parameter space is an $n$-dimensional manifold $\cM$ parametrized locally by coordinates $\lambda^i$. In this case, $\cA_R$ and $\cA_{L^{\st}}$ become one-forms on this parameter space. It is natural to interpret them as connection one-forms for a $\boldsymbol{U}(\text{dim}\,\cH_L)\times \boldsymbol{U}(\text{dim}\,\cH_R)$ bundle over the base space $\cM$, where $\boldsymbol{U}(D)$ is the unitary group. To see this more explicitly, imagine that we consider a modified state $\Psi' =U \Psi$, where $U$ is a one-sided unitary transformation acting on $R$, but we can let $U$ depend on the parameters $\cl^i$. Then, it follows from a short calculation (using the defining equation \eqref{conn0R}) that the connections transform as
\beq
\cA_{L}' = \cA_{L},
\eeq
\beq
\cA_R' = U\,\cA_R \,U^{-1} - i dU\,U^{-1},
\eeq
which is precisely the transformation property of a connection 1-form. The same formula is also true for the transformation of $\cA_{L}$ under a one-sided unitary acting on $L$. Thus, $\cA_{R}$ and $\cA_{L}$ are connection 1-forms under the action of local, one-sided unitary transformations, and we can think of equation \eqref{transport} as defining transport with respect to these connections. We will refer to these connections as \emph{modular Berry connections}. The curvature for these connections must only lie along the diagonal $U(1)^{\mathrm{dim}\,\cH}$ subgroups in the non-degenerate case. However, the curvature is much more interesting to study in the degenerate case, where one encounters further ambiguities in how to transport eigenstates within degenerate subspaces; this is a non-Abelian generalization of the phase ambiguities we encountered previously (see \cite{Czech:2017zfq, Czech:2018kvg, Czech:2019vih} for some related work on modular Berry connections).

Coming back to the case with one parameter $\cl$, the general solution to the differential equation \eqref{diff} takes the form:\footnote{The flow equation satisfied -- for instance, by $U_R$ -- is a regulated version of the flow equation satisfied by the Connes cocycle $u_s = e^{is K_R^{(\cl)}}e^{-isK_R^{(0)}}$, in the large $s$ limit; see \cite{Ceyhan:2018zfg, Lashkari:2019ixo, Bousso:2020yxi, Levine:2020upy} for some recent discussions of the Connes cocycle.}
\beq
\cU_{\lambda} = U^{\st}_{L}(\cl)\cdot \cU_0 \cdot U_R^{\dagger}(\cl),
\eeq
where 
\beq \label{udiff2}
i\frac{dU^{\st}_{L}}{d\cl} = \cA^{\st}_{L} U^{\st}_{L},\;\;\; -i\frac{dU_R}{d\cl}= \cA_RU_{R},\;\;\;U^{\st}_{L}(0) = \mathbb{1}_{L^{\st}},\;\; U_R(0) = \mathbb{1}_R,
\eeq
The formal solutions to these equations are given by
\beq
U^{\st}_{L} = \mathcal{P}\exp\left\{-i\int_0^{\cl}d\cl'\cA^{\st}_{L}(\cl')\right\},\;\;\;U_R = \mathcal{P}\exp\left\{i\int_0^{\cl}d\cl'\cA_R(\cl')\right\},
\eeq
where $\mathcal{P}$ stands for path-ordering. The matrices $U_R$ and $U_{L}$ supply a notion of parallel transport. This, in principle, allows us to completely solve for the reflection operator $\cU_{\cl}$ in terms of the modular Hamiltonians of the left and right subregions for the one-parameter family of states $\Psi_{\cl}$.\footnote{Note that the reflection operator only depends on $a_L$ and $a_R$ through the combination $(a_L + a_R)$. We also need to impose the Schmidt condition to fix this phase ambiguity, as discussed previously. With this, the reflection operator is completely determined, but this phase ambiguity will not be important for us.}

\subsection{Expanding around the TFD state}
So far, we have derived a general differential equation satisfied by the operator $\cU_{\cl}$ for a one-parameter family of states $\Psi_{\cl}$. Now we wish to apply this to a more concrete setting. Let us consider the TFD state
\beq
|\Psi_0\rangle = \frac{1}{\sqrt{Z}}\sum_n e^{-\frac{\beta}{2}E_n(0)}|\ch_n(0)\rangle_{L}\otimes |\ch^{\st}_n(0)\rangle_R,
\eeq
where $E_n(0)$ and $\ch_n(0)$ are the eigenstates of some local Hamiltonian $H$, and the right Hilbert space $\cH_R$ can be identified with $ \cH_{L^{\st}}$. The TFD state can also be thought of as a Euclidean path integral over a Euclidean time segment of length $\beta/2$. The TFD state is itself the canonical purification of the thermal ensemble, and so the reflection operator in the present case is essentially the identity operator. 

We wish to consider a one-parameter deformation of the TFD state. A natural such family of states can be constructed by turning on a source $\tilde{J}(\tau)$ for some operator $\cO(\tau)$ in the Euclidean path integral \cite{Marolf:2017kvq}. Concretely, we change the action inside the Euclidean path integral in the following way:\footnote{We are only displaying the time coordinate here and in what follows, but in principle the sources can also depend on spatial directions.}
\beq
S_{\text{new}} = S_{\text{old}}+\cl \int_{-\pi}^0 d\tau \tilde{J}(\tau)\cO(\tau),
\eeq
where we have defined $\tau=\frac{2\pi}{\beta}\hat{\tau}$ and $\hat{\tau}$ is the Euclidean time coordinate with period $\beta$. This new path integral now constructs a new bi-partite state which we will call $\Psi_{\cl}$. We wish to construct the reflection operator $\cU_{\cl}$ for this family of states to first order in $\cl$. 

In order to do this, we first need to compute the change in the modular Hamiltonians of the $L$  and $R$ subsystems to first order in perturbation theory. This has been computed previously in several works, see for instance \cite{Faulkner:2016mzt, Sarosi:2017rsq, Lashkari:2018tjh, Balakrishnan:2020lbp}:
\beq\label{ModHam}
\frac{dK_R}{d\cl} = \int_0^{2\pi} d\tau\, J_R(\tau)\int_{-\infty}^{\infty}\frac{ds}{4\sinh^2(\frac{s+i\tau}{2})}e^{\frac{is}{2\pi}K_{R}(0)}\cO(0)e^{-\frac{is}{2\pi}K_{R}(0)}.
\eeq
Here $K_R(0)=\beta H$ is the original, undeformed modular Hamiltonian for $\Psi_0$, which is simply $\beta$ times the Hamiltonian $H$ corresponding to the TFD state. The source $J_R(\tau)$ is a time-reflection symmetric version of $\tilde{J}(\tau)$:
\beq
J_R(t) = \begin{cases} \tilde{J}(\tau) & -\pi < \tau < 0\\
\tilde{J}^*(- \tau) & 0 < \tau < \pi. \end{cases}
\eeq
Note that the operator on the right hand side of equation \eqref{ModHam} is a fully Lorentzian operator; all the Euclidean time dependence is now in the $\sinh^{-2}(\frac{s+i\tau}{2})$ kernel. A similar formula can also be written for the left subsystem. The only difference is that the corresponding source $J_L$ is related to $J_R$ by a left-right reflection, i.e., $J_L(\tau) = J_R(\pi - \tau)$. 

Let us briefly recap where equation \eqref{ModHam} comes from. In the finite dimensional case, one proceeds as follows:\footnote{We will temporarily drop the subscripts $L$ and $R$, since this derivation applies to both and the subscript is not so relevant.}
\beqn
\frac{dK}{d\cl}\Big|_{\cl= 0} &=& -\lim_{\epsilon\to 0}\frac{1}{\epsilon}\left(\log(\rho_0 +\epsilon \frac{d\rho}{d\cl})-\log \rho_0\right)\nonumber\\
&=&-\lim_{\epsilon\to 0}\frac{1}{\epsilon}\left(\log[\rho_0(1 +\epsilon \rho_0^{-1} \frac{d\rho}{d\cl})]-\log \rho_0\right)\nonumber\\
&=&-\lim_{\epsilon\to 0}\frac{1}{\epsilon}\left(\log[e^{-K(0)}e^{\epsilon \rho_0^{-1} \frac{d\rho}{d\cl}}]-\log \rho_0\right).
\eeqn
Here, we have only assumed that $\rho_0$ is invertible. Using the Baker-Campbell-Hausdorff formula in the first term, we get
\beq
\frac{dK}{d\cl}\Big|_{\cl= 0} = -\sum_{n=0}^{\infty}(-1)^n\frac{B_n}{n!}\left[K(0),\cdots\left[K(0),\rho_0^{-1} \frac{d\rho}{d\cl}\right]\cdots\right].
\eeq
Now, using the integral formula
\beq
B_n = \int_{-\infty+i\epsilon}^{\infty+i\epsilon}ds \frac{\left(\frac{-is}{2\pi}\right)^n}{4\sinh^2(s/2)},
\eeq
we can re-sum the BCH expansion to obtain
\beq
\frac{dK}{d\cl}\Big|_{\cl= 0} = -\int_{-\infty+i\epsilon}^{\infty+i\epsilon}\frac{ds}{4\sinh^2(s/2)} e^{\frac{is}{2\pi}K(0)}\rho_0^{-1} \frac{d\rho}{d\cl}e^{-\frac{is}{2\pi}K(0)}.
\eeq
For Euclidean path-integral states, a path-integral argument \cite{Rosenhaus:2014zza} shows that\footnote{More precisely, the operator which appears in this equation is $:\cO:=\cO-\langle \cO\rangle_0$, but for simplicity, we can assume that the one point function of $\cO$ vanishes.}
\beq
\rho_0^{-1} \frac{d\rho}{d\cl} = -\int_0^{2\pi}d\tau J(\tau)\cO(\tau),
\eeq
so we obtain
\beq
\frac{dK}{d\cl}\Big|_{\cl= 0} = \int d\tau J(\tau)\int_{-\infty+i\epsilon}^{\infty+i\epsilon}\frac{ds}{4\sinh^2(s/2)} e^{\frac{is}{2\pi}K(0)}\cO(\tau)e^{-\frac{is}{2\pi}K(0)}.
\eeq
In the finite-dimensional case, this expression is good enough, but we would like to obtain a formula which is well-defined in the infinite dimensional or continuum quantum field theory limit as well. In the latter case, the above expression becomes problematic, since the operator $\cO(\tau)$ is a Euclidean operator and does not admit a bounded continuum limit for all $\tau$. In order to avoid this problem, we first deform the $s$-contour integral\footnote{The integrand is analytic in the $0<\text{Im}(s)<2\pi$ strip of the complex $s$-plane. Furthermore, in the finite dimensional setting, the vertical contours at $s=\pm \infty$ can be dropped because $\sinh^{-2}(s/2)$ decays exponentially.} (before taking the continuum limit), to write the above expression as 
\beq
\frac{dK}{d\cl}\Big|_{\cl= 0} = \int d\tau J(\tau)\int_{-\infty}^{\infty}\frac{ds}{4\sinh^2(\frac{s+i\tau}{2})} e^{\frac{is}{2\pi}K(0)}\cO(0)e^{-\frac{is}{2\pi}K(0)}.
\eeq
Now we have a completely Lorentzian operator at hand, and at this stage we can take the continuum limit to obtain a well-defined continuum operator. 

With the first order change of the modular Hamiltonian in hand, we can now obtain the first order change in $U_R$:
\beq
-i\frac{dU_R}{d\cl}(0) = \cA_R(0),
\eeq
where
\beq
\cA_R(0) = a_R(0)+\int_0^{\infty}dt\,e^{-\epsilon t}\int d\tau J_R(\tau)\int_{-\infty}^{\infty}\frac{ds}{4\sinh^2(\frac{s+i\tau}{2})}e^{\frac{i(s+2\pi t)}{2\pi}K_{R}(0)}\cO(0)e^{-\frac{i(s+2\pi t)}{2\pi}K_{R}(0)}.
\eeq
Shifting $s$ by $2\pi t$ allows us to perform the $t$ integral:
\beq
\int_0^{\infty}idt\,\frac{e^{-\epsilon t}}{4\sinh^2(\frac{s-2\pi t+i\tau}{2})} = \frac{1}{2\pi i}\frac{1}{\left(1-e^{-(s+i\tau)}\right)}+\frac{\epsilon}{\pi^2}e^{-\frac{\epsilon}{2\pi}(s+i\tau)}B_{e^{s+i\tau}}(1+\frac{\epsilon}{2\pi},0)
\eeq
where 
$$B_z(a,b)= \int_0^z dt\,t^{a-1}(1-t)^{b-1},$$
is the incomplete Beta function. In the $\epsilon\to 0$ limit, the second term drops out, as long as the source $J_R(\tau)$ is supported away from $\tau = 0$. Thus, we get
\beq \label{MBC1}
\cA_R(0) =a_R(0)- \frac{1}{2\pi }\int d\tau J_R(\tau)\int_{-\infty}^{\infty} ds\frac{1}{\left(1-e^{-(s+i\tau)}\right)}e^{\frac{is}{2\pi}K_{R}(0)}\cO(0)e^{-\frac{is}{2\pi}K_{R}(0)}.
\eeq
Similarly, 
\beq\label{MBC2}
\cA_L(0) = a_L(0)- \frac{1}{2\pi }\int d\tau J_L(\tau)\int_{-\infty}^{\infty} ds\frac{1}{\left(1-e^{-(s+i\tau)}\right)}e^{\frac{is}{2\pi}K_{L}(0)}\cO(0)e^{-\frac{is}{2\pi}K_{L}(0)}.
\eeq

Equations \eqref{MBC1} and \eqref{MBC2} are our main formulas for the modular Berry connections evaluated on the TFD state. In the next section, we will use these to derive the quantum extremal shock in the Engelhardt-Wall geometry. As another application of these formulas, it is not hard to show that in holographic conformal field theories, these expressions for the modular Berry connections can be put in a manifestly geometric form in the bulk. Indeed, when $\cO$ is taken to be a single-trace operator, we find that
\beq\label{SF1}
\Pi_{\text{code}}\cA_R(0)\Pi_{\text{code}} = \Pi_{\text{code}}a_R(0)\Pi_{\text{code}} + \int_{\Sigma_R} \bs{\omega}(\delta_{\cl}\phi,\boldsymbol{\phi}),
\eeq
\beq\label{SF2}
\Pi_{\text{code}}\cA_L(0)\Pi_{\text{code}} = \Pi_{\text{code}}a_L(0)\Pi_{\text{code}} + \int_{\Sigma_L} \bs{\omega}(\delta_{\cl}\phi,\bs{\phi}).
\eeq
Here, $\Pi_{\text{code}}$ is the projector onto states where we can think of the bulk in terms of quantum fields on a fixed background geometry, $\bs{\phi}$ is the bulk operator valued field dual to $\cO$, $\delta_{\cl}\phi$ is the linearized change in the bulk field configuration under the boundary deformation $J_R$, and $\bs{\omega}$ is the symplectic current for the bulk fields:\footnote{In the case where $\cO$ is the stress tensor, one uses the gravitational symplectic form which appears naturally in the covariant phase space method \cite{Iyer:1994ys}. The region of integration for the gravitational symplectic flux turns out to be the entanglement wedge of the boundary subregion in the deformed geometry. }
\beq
\bs{\omega}(\delta_1\phi, \delta_2\phi) = (\delta_1\phi\, n^{\mu}\pa_{\mu}\delta_2\phi - \delta_2\phi\, n^{\mu}\pa_{\mu}\delta_1\phi).
\eeq
The derivation of equations \eqref{SF1} and \eqref{SF2} more or less follows the same logic as in \cite{Faulkner:2017tkh}, so we will not repeat it here. These equations give a natural generalization of \cite{Belin:2018fxe} to the case of subregions (see also \cite{Kirklin:2019ror} for a different approach). It is intriguing that the above expressions can be written as a sum of two terms, where the first term comes from the ``diagonal'' part of the connection, while the second term is related to the symplectic flux of bulk quantum fields in the relevant entanglement wedge; it would be interesting to understand the first term better. One thing to note is that if the source $J_R$ is tuned in order to create a localized excitation at some point in the bulk, then the geometric term in $\cA_R$ is also localized at that point. Thus, the deeper in the bulk the excitation created by the source, the more ``complex'' is the corresponding unitary $U_R$. 
%\subsection{Second order}
%\subsection{General formula at all orders}
\section{Quantum extremal shock}\label{sec:gravity}

In this section, we wish to study the state of bulk matter in the holographic dual corresponding to the canonical purification $\Psi_{\cl}^{\st}$. To be concrete, we will work to first order in perturbation theory near the TFD state.

\subsection{Double-trace deformation}
We wish to turn on an operator $\cO$ in the Euclidean path-integral which sources the bulk stress tensor at $O(\cl)$. The reason is that in order to see the quantum extremal shock at $O(\cl)$ in the canonically purified state, we need to have a non-trivial shape derivative for the bulk entanglement entropy at $O(\cl)$ in the original state. But to linear order in $\cl$, we have
\beqn \label{EEfromANEC}
\frac{1}{\sqrt{h(y^i)}}\frac{d}{d\cl}\frac{\delta S_{\text{bulk}}}{\delta x^+}\Big|_{\cl=0,y^i} &=&  - 2\pi\int_0^{\infty} dx^+ \frac{d}{d\cl}\langle T_{++}^{\text{bulk}}(x^+,x^-=0,y^i)\rangle_{\Psi_{\cl}}\Big|_{\cl=0},\nonumber\\
&=& 2\pi\int^0_{-\infty} dx^+ \frac{d}{d\cl}\langle T_{++}^{\text{bulk}}(x^+,x^-=0,y^i)\rangle_{\Psi_{\cl}}\Big|_{\cl=0},
\eeqn
with a similar equation for the shape derivative along $x^-$. Here $(x^+,x^-)$ are light-cone coordinates on which Schwarzschild boosts act simply as $(x^+,x^-) \to (x^+e^s,x^-e^{-s})$, $y^i$ are transverse bulk coordinates which parametrize the original extremal surface (i.e., the bifurcation point), $h$ is the determinant of the induced metric on the original extremal surface, and the shape derivative at the point $y^i$ is defined as
\beq
\frac{\delta S_{\text{bulk}}}{\delta x^+}\Big|_{\cl=0,y^i} = \lim_{\epsilon \to 0}\frac{1}{\epsilon}\left[S_{\text{bulk}}[x^+=\epsilon \delta(y^i),x^-=0] -S_{\text{bulk}}[x^+=0,x^-=0]\right],
\eeq
where the arguments of the entropies on the right hand side above are the coordinate locations of the corresponding bulk entanglement cuts. We can derive equation \eqref{EEfromANEC} as follows: consider the bulk relative entropy for the region corresponding to the entanglement wedge $r$ of the boundary subregion $R$:
\beq
S_{\text{bulk}}(\rho_{r}(\cl)||\rho_{r}(0)) = \Delta \langle K_{\text{bulk},r}(0)\rangle - \Delta S_{\text{bulk}},
\eeq
where the $\Delta$ symbol stands for subtraction with respect to the background TFD state:
\beq
\Delta \langle K_{\text{bulk},r}(0)\rangle = \langle K_{\text{bulk},r}(0)\rangle_{\Psi_{\cl}}-\langle K_{\text{bulk},r}(0)\rangle_{\Psi_0},
\eeq
\beq
\Delta S_{\text{bulk}} = S_{\text{bulk}}(\Psi_{\cl})- S_{\text{bulk}}(\Psi_{0}).
\eeq
Since the first derivative of the relative entropy at $\cl=0$ vanishes, we conclude that
\beq
\frac{d}{d\cl}S_{\text{bulk}}\Big|_{\cl=0}= \frac{d}{d\cl}\langle K_{\text{bulk},r}(0)\rangle_{\Psi_{\cl}}\Big|_{\cl=0}.
\eeq
Taking a derivative of this equation with respect to the shape of the bulk entanglement cut and using \cite{Faulkner:2016mzt, Casini:2017roe}
\beq
\frac{\delta K_{\text{bulk},r}(0)}{\delta x^+}\Big|_{y^i} = -2\pi\sqrt{h(y^i)}\int_0^{\infty} dx^+  T^{\text{bulk}}_{++}(x^+,x^-=0,y^i),
\eeq
we land on the first equality in equation \eqref{EEfromANEC}, while applying the same arguments to the the entanglement wedge $\ell$ of $L$ and using 
\beq
\frac{\delta K_{\text{bulk},\ell}(0)}{\delta x^+}\Big|_{y^i} = 2\pi\sqrt{h(y^i)}\int^0_{-\infty} dx^+  T^{\text{bulk}}_{++}(x^+,x^-=0,y^i),
\eeq
gives the second equality. Importantly, equation \eqref{EEfromANEC} implies that for us to see the shock in the bulk dual to the canonical purification at $O(\cl)$, we need to turn on a deformation which will source the bulk stress tensor at $O(\cl)$ in the original state. For this reason, we cannot take $\cO$ to be a single-trace operator, as single trace operators only source the bulk stress tensor at $O(\cl^2)$. Instead, we can imagine turning on a double-trace operator $\cO =\;:\phi\phi:$, for some single trace operator $\phi$; although the details of what $\cO$ we choose will not be relevant in the discussion below. Now, the quantum extremal surface in the geometry dual to $\Psi_{\cl}$ will deviate from the classical extremal surface at $O(\cl G_N)$. Following the Engelhardt-Wall construction reviewed in section \ref{sec:EWconstruction}, the bulk spacetime dual to the canonical purification $\Psi_{\cl}^{\st}$ consists of the entanglement wedge $\text{EW}(L)$ (in the original geometry dual to $\Psi_{\cl}$) glued to its CPT image at the QES. In order for the junction conditions to be satisfied, the bulk matter stress tensor must have a singular contribution at the location of the QES. Importantly, even though the QES deviates from the classical extremal surface at $O(\cl G_N)$, the singular contribution in the bulk stress tensor in the bulk dual to the canonically purified state must appear at $O(\cl)$. It is this contribution that we are after.  

\subsection{Bulk one point function}
In order to proceed, we wish to compute the bulk stress tensor in the canonically-purified state. We can be general, and compute the one-point function of a more general operator $\Phi$ acting on the $L^{\st}$ factor:
\beq \label{1pt}
\langle \Phi \rangle_{\Psi_{\cl}^{\st}} = \langle \Psi_{\lambda} | \cU^{\dagger}_{\lambda}\,\Phi\, \cU_{\lambda} |\Psi_{\lambda}\rangle,
\eeq
at first order in $\lambda$. Later, we will take $\Phi$ to be the bulk matter stress tensor $T^{\text{bulk}}_{\mu\nu}(x_B)$, where we will take $x_B$ to lie in the entanglement wedge of $L^{\st}$ in the geometry dual to $\Psi_{\cl}^{\st}$. In particular, we are interested in the $T^{\text{bulk}}_{\pm\pm}$ components of the stress tensor, and we wish to take the limit where the bulk point approaches the quantum extremal surface. Let us take a moment to discuss what this means. The backreaction from turning on a double-trace operator is of $O(\lambda G_N)$. If we ignore this effect for now, the classical bulk spacetime dual to the canonically purified state is the undeformed, eternal black hole spacetime, where we simply re-label the right subsystem as $L^{\st}$. However, the state of bulk matter fields receives corrections at $O(\lambda)$, and this is what we wish to probe via the bulk operator $\Phi$; in particular, we want to take $\Phi = T^{\text{bulk}}_{\pm\pm}$ and take the limit where this operator approaches the original extremal surface (i.e., the bifurcation point) in the eternal black hole. 

With this preamble, we now wish to compute the first $\cl$ derivative of the above one-point function. Taking a $\cl$-derivative of equation \eqref{1pt}, we get:
\beq
\frac{d}{d\cl}\langle \Phi \rangle_{\lambda,\st} = \langle \Psi_{\cl}|\frac{d\cU_{\cl}^{\dagger}}{d\cl}\,\Phi\, \cU_{\lambda}|\Psi_{\lambda}\rangle+ \langle \Psi_{\cl}|\cU_{\cl}^{\dagger}\Phi\, \frac{d\cU_{\cl}}{d\cl}|\Psi_{\lambda}\rangle + \langle \frac{d\Psi_{\cl}}{d\cl}|\widehat{\Phi}|\Psi_{\cl}\rangle + \langle \Psi_{\cl}|\widehat{\Phi}|\frac{d\Psi_{\cl}}{d\cl}\rangle,
\eeq
where in the last two terms we have defined the operator $\widehat{\Phi} \equiv\cU_{\cl}^{\dagger}\,\Phi\,\cU_{\cl}$. Using the flow equation for $\cU_{\cl}$, we can rewrite this as
%&=& \langle \Psi_{\cl}|\left(-\cU^{\dagger}_{\cl}\cA_L+\cA_R\cU^{\dagger}_{\cl}\right)\Phi\, \cU_{\lambda}|\Psi_{\lambda}\rangle+ \langle \Psi_{\cl}|\cU_{\cl}^{\dagger}\,\Phi \left(\cA_L\cU_{\cl}-\cU_{\cl}\cA_R\right)|\Psi_{\lambda}\rangle\nonumber\\
%&=&\langle \Psi_{\cl}|\left[\cA_R, \cU_{\cl}^{\dagger}\,\Phi\,\cU_{\cl}\right]|\Psi_{\cl}\rangle-\langle \Psi_{\cl}|\cU_{\cl}^{\dagger}\left[\cA_{L^{\st}},\Phi\right]\cU_{\cl}|\Psi_{\cl}\rangle\nonumber\\
\beq \label{3terms}
\frac{d}{d\cl}\langle \Phi \rangle_{\lambda,\st}=i\langle \Psi_{\cl}|\left[\cA_R, \widehat{\Phi}\right]|\Psi_{\cl}\rangle-i\langle \Psi^{\st}_{\cl}|\left[\cA^{\st}_{L^{\st}}, \Phi\right]|\Psi^{\st}_{\cl}\rangle+\langle \frac{d\Psi_{\cl}}{d\cl}|\widehat{\Phi}|\Psi_{\cl}\rangle + \langle \Psi_{\cl}|\widehat{\Phi}|\frac{d\Psi_{\cl}}{d\cl}\rangle.
\eeq 
Note that at $\cl=0$, $\widehat{\Phi}=\Phi$, and so henceforth we will drop the hats. Further, the last two terms can simply be written as
\beq
\left(\langle\frac{d\Psi_{\cl}}{d\cl}|\widehat{\Phi}|\Psi_{\cl}\rangle + \langle \Psi_{\cl}|\widehat{\Phi}|\frac{d\Psi_{\cl}}{d\cl}\rangle\right)_{\cl=0}=\frac{d}{d\cl}\langle \Phi\rangle_{\Psi_{\cl}}\Big|_{\cl=0}.
\eeq

Let us now focus on the first term involving the commutator with $\cA_R$; the same logic will also apply to the second term. We proceed by assuming that $\Phi(x_B)$ is an operator acting strictly on the $\cH_{R}$ factor (i.e., $x_B$ is well within the entanglement wedge of $R$). As explained above, we will eventually take $\Phi = T^{\text{bulk}_{\pm\pm}}$ and take the limit where the bulk point approaches the bifurcation point. To be precise, when the operator acts ``at the bifurcation point'', we cannot take it to be supported in $\cH_R$ alone. For instance, after a little bit of smearing to make this bulk operator well-defined, we will in general find that it acts on both sides of the bifurcation surface. A simple smearing is to instead consider the operators 
\beq\label{smear}
\Phi_{\text{smear}}= \lim_{\delta \to 0}\int_{-\delta}^{\delta}dx^{\pm}\,T^{\text{bulk}}_{\pm\pm}.
\eeq
Indeed, later we will encounter the need for such a smearing, but for now we proceed with the above simplifying assumption. If we work at $\cl=0$, and use equation \eqref{MBC1}:
\beq 
\cA_R(0) =a_R(0)- \frac{1}{2\pi }\int d\tau J_R(\tau)\int_{-\infty}^{\infty} ds\frac{1}{\left(1-e^{-(s+i\tau)}\right)}e^{\frac{is}{2\pi}K_{R}(0)}\cO(0)e^{-\frac{is}{2\pi}K_{R}(0)}.
\eeq
then we get
\beqn
\langle \Psi_{0}|\left[\cA_{R}(0), \Phi(x_B)\right]|\Psi_{0}\rangle &=& \mathrm{Tr}_{R}\left(\rho^{(0)}_{R}\left[\cA_{R}(0), \Phi(x_B)\right]\right)\\
&=&\frac{1}{2\pi i}\int d\tau J_R(\tau)\int_{-\infty}^{\infty}\frac{ds}{\left(1 - e^{-(s+i\tau)}\right)}\,\mathrm{Tr}_{R}\left(\rho^{(0)}_{R}\left[\cO(s),\Phi\right]\right)\nonumber\\
&=&\frac{1}{2\pi i}\int d\tau J_R(\tau)\int_{-\infty-i\epsilon}^{\infty-i\epsilon}\frac{ds}{\left(1 - e^{-(s+i\tau)}\right)}\,\mathrm{Tr}_{R}\left(\rho_{R}^{(0)}\cO(s)\Phi\right)\nonumber\\
&-&\frac{1}{2\pi i}\int d\tau J_R(\tau)\int_{-\infty-i(2\pi-\epsilon)}^{\infty-i(2\pi-\epsilon)}\frac{ds}{\left(1 - e^{-(s+i\tau)}\right)}\,\mathrm{Tr}_{R}\left(\rho_{R}^{(0)}\cO(s)\Phi\right).\nonumber
\eeqn
\begin{figure}
    \centering
    \includegraphics[height=4cm]{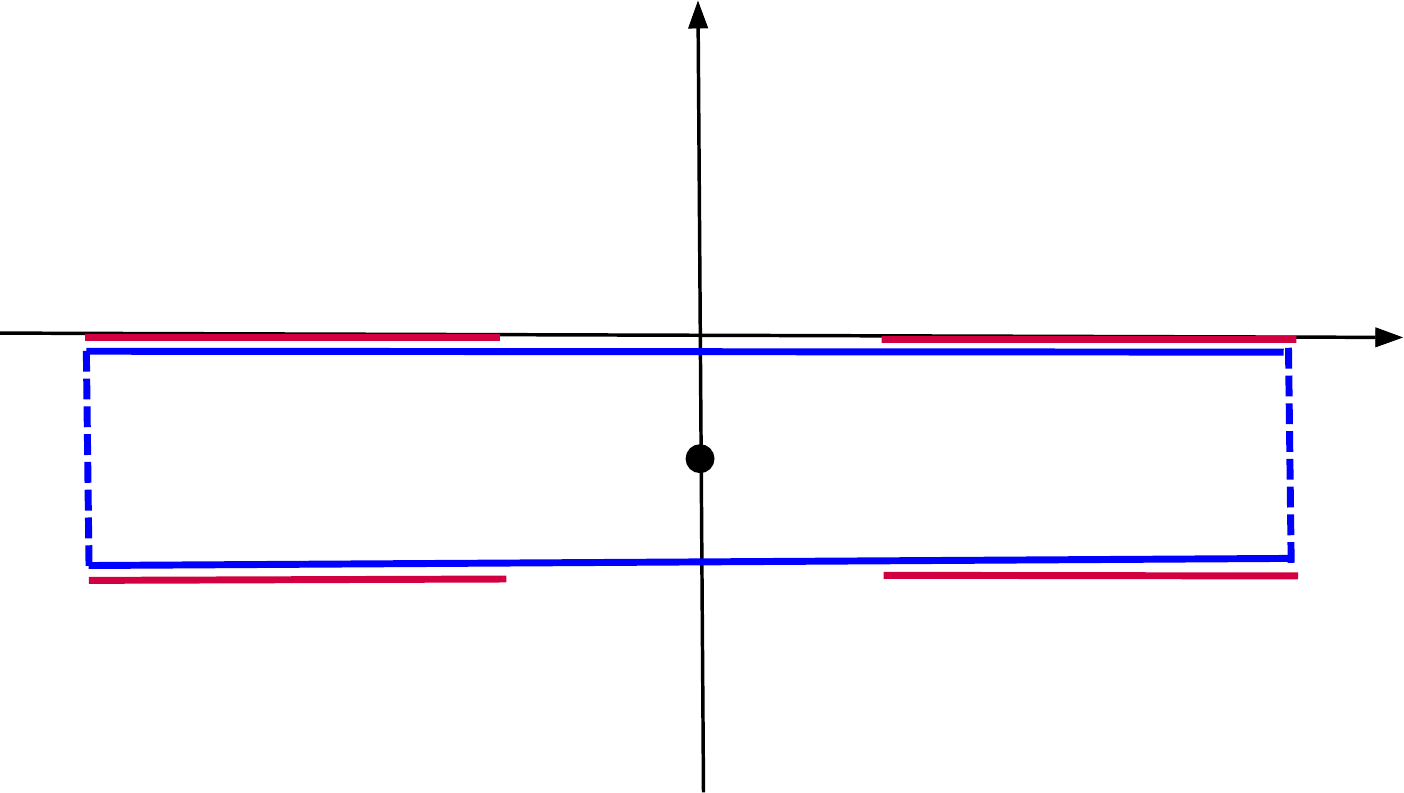}
    \caption{The strip $-2\pi \leq \text{Im}(s)\leq 0$ in the complex-$s$ plane. The contour $\Gamma$ is shown in the bold blue. The dashed blue lines indicate the vertical contours at infinity. The red lines indicate potential branch cuts which may develop in the correlation function in the infinite dimensional limit, while the black dot indicates the pole coming from $\frac{1}{1-e^{-(s+i\tau)}}$.}
    \label{fig:contour}
\end{figure}
In the second line, we have used the fact that $a_R(0)$ commutes with $\rho_R^{(0)}$ to drop that term. In the third line, we have introduced a new regulator $\epsilon \to 0^+$ to separate the two operators infinitesimally in Euclidean time, and further used the KMS condition to bring the two operators in the same order. So, we conclude that
\beq\label{corfunc}
\langle \Psi_{0}|\left[\cA_{R}, \Phi(x_B)\right]|\Psi_{0}\rangle = \frac{1}{2\pi i}\int d\tau\,J_R(\tau)\int_{\Gamma}\frac{ds}{\left(1 - e^{-(s+i\tau)}\right)}\,\mathrm{Tr}_{R}\left(\rho_{R}^{(0)}\cO(s)\Phi\right),
\eeq
where the contour $\Gamma = (\mathbb{R}-i\epsilon) \cup (\mathbb{R}-i(2\pi - \epsilon))$ is the union of the two horizontal contours at $\text{Im}(s)=-\epsilon$ and $\text{Im}(s) = - (2\pi -\epsilon)$.

Using Cauchy's theorem, we can then rewrite this integral as the sum over three contributions: the pole at $s=-i\tau$, and the two ``vertical'' contours at $\mathbb{Re}(s) = \pm \Lambda$ (with $\Lambda \to \infty$):
\beq
\langle \Psi_{0}|\left[\cA_{R}(0), \Phi(x_B)\right]|\Psi_{0}\rangle = -\int d\tau\,J_R(\tau)\mathrm{Tr}_{R}\left(\rho^{(0)}_{R}\cO(\tau)\Phi\right)+\mathcal{I}^R_++\mathcal{I}^R_-,
\eeq
where
\beq\label{VC1}
\mathcal{I}^R_{\pm}=\pm \frac{1}{2\pi}\int d\tau\,J_R(\tau)\int_{\epsilon}^{2\pi-\epsilon}\frac{d\theta}{\left(1 - e^{-(\pm\Lambda+i\tau)}e^{i\theta}\right)}\,\mathrm{Tr}_{R}\left(\rho^{(0)}_{R}e^{i(\pm\Lambda-i\theta)K_{R}(0)}\cO e^{-i(\pm\Lambda-i\theta)K_{R}(0)}\Phi\right).
\eeq
The vertical contour contributions are seemingly suppressed exponentially in $\Lambda$ from the large relative boost between the two operators, and so it is tempting to discard them. This is correct in most cases, but not all; we will return to this point below, where we will find that the shocks we are looking for actually come from these terms. For now, let us focus on the contribution of the pole: 
\beqn
\langle \Psi_{0}|\left[\cA_{R}(0), \Phi(x_B)\right]|\Psi_{0}\rangle\Big|_{\text{pole}} &=& -\int d\tau\,J_R(\tau)\mathrm{Tr}_{R}\left(\rho^{(0)}_{R}\cO(\tau)\Phi\right)\nonumber\\
&=&-\int d\tau\,J_R(\tau)\left\langle \cO(\tau)\Phi\right\rangle_{\Psi_0}\nonumber\\
&=&-\frac{d}{d\cl}\langle\Phi\rangle_{\Psi_\cl}\Big|_{\cl=0}.
\eeqn
This term simply cancels the last term in equation \eqref{3terms}. This is expected: this term measures how the entanglement wedge of $R$ would change in the geometry dual to $\Psi_{\cl}$, but in the canonical purification, the entanglement wedge of $R$ is replaced by a CPT reflected image of the entanglement wedge of $L$. Thus, the above cancellation ensures that all information about the entanglement wedge of $R$ is removed. We must now show that, in fact, the entanglement wedge of $R$ is replaced with a CPT image of the entanglement wedge of $L$. This comes from the pole contribution in the second term of \eqref{3terms}:
\beqn
\langle \Psi^{\star}_{0}|\left[\cA^{\st}_{L^{\st}}(0), \Phi(x_B)\right]|\Psi^{\st}_{0}\rangle\Big|_{\text{pole}} &=& \int d\tau\,J_L(\tau)\mathrm{Tr}_{L^{\st}}\left(\rho^{(0)}_{L^{\st}}\Theta\,\cO(\tau)\,\Theta^{-1}\,\Phi\right)\nonumber\\
&=&\int d\tau\,J_L(\tau)\mathrm{Tr}_{L}\left(\rho^{(0)}_{L}\cO(\tau)\,\Theta^{-1}\,\Phi\,\Theta\right)\nonumber\\
&=&\frac{d}{d\cl}\langle\Phi^{\st}\rangle_{\cl=0},
\eeqn
where $\Phi^{\st}= \Theta^{-1}\Phi \Theta$ is the CPT conjugate of the operator, but now inserted in the entanglement wedge of $L$. This is in precise agreement with our expectation for what the canonical purification should do. 
\subsection{Vertical contours at infinity}
So far, we have reproduced the standard, expected properties of the bulk dual to the canonical purification. Now we turn to the non-trivial part, which is to reproduce the quantum extremal shock in the bulk. For this, we need to choose a specific bulk operator, i.e., we need to take $\Phi = T^{\text{bulk}}_{\pm\pm}$. Moreover, we need to take the limit where this bulk operator approaches the extremal surface in the original background geometry. To be concrete, let us take $\Phi=T^{\text{bulk}}_{++}$ and consider the vertical contour integral $\mathcal{I}_{\pm}^R$:
\beq
\mathcal{I}^R_{\pm} =\pm \frac{1}{2\pi}\int d\tau\,J_R(\tau)\int_{\epsilon}^{2\pi-\epsilon}\frac{d\theta}{\left(1 - e^{-(\pm\Lambda+i\tau)}e^{i\theta}\right)}\,\mathrm{Tr}_{R}\left(\rho_{R}^{(0)}\cO e^{i(\mp\Lambda+i\theta)K_{R}(0)}T^{\text{bulk}}_{++}e^{-i(\mp\Lambda+i\theta)K_{R}(0)}\right)
\eeq
This is the same as \eqref{VC1}, but we have now put the boost on the bulk operator. As $\Lambda \to \infty$, the relative boost between the two operators goes off to infinity, and so we expect the correlator to decay exponentially. Thus, in the $\Lambda \to \infty$ limit, this contour integral vanishes. The exception to this occurs when the bulk operator approaches the extremal surface. To see this, let us use light-cone coordinates $(x^+,x^-)$ in the plane transverse to the black hole extremal surface. At $\cl=0$, boundary modular flow acts locally on bulk operators as a Schwarzschild boost:
\beq
e^{is K_{R}(0)}T^{\text{bulk}}_{\mu\nu}(x^+,x^-,y^i)e^{-isK_{R}(0)}={J^{\alpha}}_{\mu}(s){J^{\beta}}_{\nu}(s) T^{\text{bulk}}_{\alpha\beta}(x^+e^{s} ,x^-e^{-s},y^i),
\eeq
where the ${J^{\alpha}}_{\beta}$ represents the action of the boost on the indices of the stress tensor. In more detail,
\beq
e^{is K_{R}(0)}T^{\text{bulk}}_{\pm\pm}(x^+,x^-,y^i)e^{-isK_{R}(0)}=e^{\pm 2s} T^{\text{bulk}}_{\pm \pm}(x^+e^{s} ,x^-e^{-s},y^i),
\eeq
\beq
e^{is K_{R}(0)}T^{\text{bulk}}_{\pm i}(x^+,x^-,y^i)e^{-isK_{R}(0)}=e^{\pm s} T^{\text{bulk}}_{\pm i}(x^+e^{s} ,x^-e^{-s},y^i),
\eeq
\beq
e^{is K_{R}(0)}T^{\text{bulk}}_{ij}(x^+,x^-,y^i)e^{-isK_{R}(0)}= T^{\text{bulk}}_{ij}(x^+e^{s} ,x^-e^{-s},y^i).
\eeq
Consider first $\mathcal{I}_-^R$:
\beqn \label{ir-}
\mathcal{I}_-^R &=& \frac{1}{2\pi}\int d\tau\,J_R(\tau)\int_{\epsilon}^{2\pi-\epsilon}\frac{d\theta e^{2\Lambda}}{\left(1 - e^{(\Lambda - i\tau)}e^{i\theta}\right)}\,\mathrm{Tr}_{R}\left(\rho_{R}^{(0)}\cO e^{-\theta K_{R}(0)}T^{\text{bulk}}_{++}(x^+e^{\Lambda}, x^- e^{-\Lambda})e^{\theta K_{R}(0)}\right),\nonumber\\
&\simeq & -\frac{1}{2\pi}\int d\tau\,J_R(\tau)\int_{\epsilon}^{2\pi-\epsilon} d\theta e^{\Lambda +i(\tau-\theta)}\,\mathrm{Tr}_{R}\left(\rho^{(0)}_{R}\cO e^{-\theta K_{R}(0)}T^{\text{bulk}}_{++}(x^+e^{\Lambda}, x^- e^{-\Lambda})e^{\theta K_{R}(0)}\right).
\eeqn
If $x^+ \neq 0$, then the above correlation function will decay exponentially in $\Lambda$ as previously mentioned, and is thus zero in the $\Lambda \to \infty$ limit because the bulk operator is getting boosted off to infinity. However, when $x^+=0$, the operator does not get boosted away, and we instead get a divergence from the $e^{\Lambda}$ factor in equation \eqref{ir-}.\footnote{A similar effect is responsible for the Ceyhan-Faulkner shock \cite{Ceyhan:2018zfg} in Connes-cocyle flowed states in the perturbative setup \cite{Bal}.} We can see this quite explicitly in the BTZ black hole, for instance. In Kruskal coordinates, the bulk metric is given by
\beq
ds^2 = -\frac{4dx^+dx^-}{(1+x^+x^-)^2}+ \frac{4\pi^2}{\beta^2}\frac{(1-x^+x^-)^2}{(1+x^+x^-)^2} d\phi^2,
\eeq
where $\phi$ is a periodic coordinate along the bifurcation surface. The bulk to boundary propagator is given by \cite{Keski-Vakkuri:1998gmz, Maldacena:2001kr}
\beq \label{BBP}
K(x^+,x^-,\phi)=\sum_{n}\frac{(1+x^+x^-)^{2\Delta}}{\left\{(1-x^+x^-)[\cosh(\frac{2\pi}{\beta}(\phi-\phi_0+2\pi n))-1]+(x^+-e^{-i\theta_0})(x^--e^{i\theta_0})\right\}^{2\Delta}},
\eeq
where $(\phi_0, \tau_0)$ label the coordinates on the boundary torus with $\tau_0$ being the Euclidean time direction and $\phi_0$ being the spatial direction. The bulk stress tensor in the presence of the boundary double-trace operator is given by
\beq
\langle T_{++}^{\text{bulk}} \cO\rangle \sim \sum_n\pa_+K_n \pa_+K_n,
\eeq
where $K_n$ is the $n$th term in the summation in equation \eqref{BBP}. For fixed $n$ and $x^+\neq 0$, the bulk stress tensor goes as $e^{-(4\Delta+2)s}$ in the large $s$ limit. However, when $x^+=0$, there is no suppression as the operator does not get boosted away and $\mathcal{I}^R_-$ diverges, because of the factor of $e^{\Lambda}$ out front in equation \eqref{ir-}; this suggests a delta-function contribution at $x^+=0$. To check this, we really need to smear the operator in the $x^+$ direction in an infinitesimally small window of $x^+\in [0,\delta]:$\footnote{We can think of this as the part of $\Phi_{\text{smear}}$ (see equation \eqref{smear}) which contributes to $[\cA_R,\Phi]$.}
\beqn
\int_0^{\delta} dx^+\,\mathcal{I}^R_- &=&-\frac{1}{2\pi}\int d\tau\,J_R(\tau)\int_{\epsilon}^{2\pi-\epsilon}d
\theta\int_0^{\delta e^\Lambda}d
\tilde{x}^+e^{i(\tau-\theta)}\,\mathrm{Tr}_{R}\left(\rho^{(0)}_{R}\cO e^{-\theta K_{R}(0)}T^{\text{bulk}}_{++}(\tilde{x}^+, x^-e^{-\Lambda})e^{\theta K_{R}(0)}\right)\nonumber\\
&\simeq &-\frac{1}{2\pi}\int d\tau\,J_R(\tau)\int_{\epsilon}^{2\pi-\epsilon}d
\theta\int_0^{\infty}d
\tilde{x}^+e^{i(\tau-\theta)}\,\mathrm{Tr}_{R}\left(\rho_{R}^{(0)}\cO e^{-\theta K_{R}(0)}T^{\text{bulk}}_{++}(\tilde{x}^+, 0)e^{\theta K_{R}(0)}\right),
\eeqn
where in the first line, we have defined a new coordinate $\tilde{x}^+ = x^+e^{\Lambda}$, and in the second line we have sent $\Lambda \to \infty$. By deforming the $\tilde{x}^+$ contour in the complex plane, we can remove all the $\theta$ dependence from the correlator, and replace it with $\tau$. Performing the $\theta$ integral then gives
\beq
\int_0^{\delta} dx^+\,\mathcal{I}_-^R = -\int d\tau\,J_R(\tau)\int_0^{\infty}d
x^+\,\mathrm{Tr}_{R}\left(\rho_{R}^{(0)}\cO(\tau)T^{\text{bulk}}_{++}(x^+, 0,y^i)\right) = \frac{1}{2\pi}\frac{d}{d \cl}\frac{\delta S_{\text{bulk}}}{ \delta x^+}\Big|_{\cl=0,y^i},
\eeq
where in the last equality we used equation \eqref{EEfromANEC}. Thus, the vertical contour precisely gives us the delta function contribution we had expected. Note that $\mathcal{I}_+^R$ does not give a delta function contribution because the enhancement factor of $e^{2\Lambda}$ is now replaced with a suppression factor of $e^{-2\Lambda}$. 

Similarly, we can evaluate the vertical contour contributions coming from the term involving $\cA^{\st}_{L^{\st}}$. In this case, the contour at $s=+\Lambda$ contributes:
\beq
\mathcal{I}^L_{+} = \frac{1}{2\pi}\int d\tau\,J_L(\tau)\int_{\epsilon}^{2\pi-\epsilon}\frac{d\theta}{\left(1 - e^{-(\Lambda+i\tau)}e^{i\theta}\right)}\,\mathrm{Tr}_{L}\left(\rho_{L}^{(0)}\cO e^{i(-\Lambda+i\theta)K_{L}(0)}(\Theta^{-1}T^{\text{bulk}}_{++}\Theta)e^{-i(-\Lambda+i\theta)K_{L}(0)}\right),
\eeq
where 
\beq 
\Theta^{-1}T^{\text{bulk}}_{++}(x^+,x^-,y^i)\Theta = T^{\text{bulk}}_{++}(-x^+,-x^-,y^i).
\eeq
The left-sided boost acts on this operator as:
\beq 
e^{-i\Lambda K_{L}(0)}T^{\text{bulk}}_{++}(-x^+,-x^-)e^{i\Lambda K_{L}(0)} = e^{2\Lambda}T^{\text{bulk}}_{++}(-x^+e^{\Lambda},-x^-e^{-\Lambda}).
\eeq
In the large $\Lambda$ limit, we can expand:
\beq 
\frac{1}{\left(1 - e^{-(\Lambda+i\tau)}e^{i\theta}\right) }= 1 +  e^{-(\Lambda+i\tau)} e^{i\theta}+\cdots. 
\eeq
The first term leads to a $e^{2\Lambda}$ divergence, but the $\theta$ integration kills this term, as can be seen by smearing in the infinitesimal interval $x^+ \in (-\delta,0)$. The first non-trivial contribution comes from the second term, which gives (following the same steps as before):
\beq 
\int_{-\delta}^0dx^+ \mathcal{I}^L_+ = -\int d\tau\int J_L(\tau) \int_{-\infty}^0 dx^+\mathrm{Tr}_{L}\left(\rho_{L}^{(0)}O(\tau)T_{++}^{\text{bulk}}(x^+,0,y^i)\right)= -\frac{1}{2\pi}\frac{d}{d\cl}\frac{\delta S_{\text{bulk}}}{\delta x^+}\Big|_{\cl=0,y^i}.
\eeq
The extra minus sign above cancels with the minus sign in front of the $\mathcal{A}_{L^\st}$ term, and thus we get the same vertical contribution from here as we had from the $\mathcal{A}_R$ term, resulting in an overall factor of 2. Thus, we learn that the bulk stress tensor has the following shock contribution in the canonically purified state:
\beq
2\pi\frac{d}{d\cl}\langle T_{++}^{\text{bulk}}(x^+,x^-,y^i)\rangle_{\Psi^{\st}_{\cl}}\Big|_{\cl=0} = 2\delta(x^+) \frac{d}{d\cl} \frac{\delta S_{\text{bulk}}}{\delta x^+}\Big|_{\cl=0,y^i}+\cdots,
\eeq
where the $\cdots$ indicate the other non-singular parts. This is precisely the shock required to support the Engelhardt Wall geometry.\footnote{Our calculation is valid in the limit $x^+\to 0$ with $x^-$ fixed. However, we see that the dependence on $x^-$ is trivial in the end. This is a simple consequence of the conservation of the shock stress tensor, $\pa_-T^{\text{shock}}_{++}=0$.} Thus, the boundary entanglement structure in the canonically purified state gives rise to a state of the matter fields in the bulk which precisely supports the Engelhardt-Wall geometry, in a way consistent with the bulk Einstein's equations.

\section{Discussion}\label{sec:discussion}
To summarize, we have studied the canonical purification of Euclidean path integral states to first order in sources. In holographic conformal field theories, we have demonstrated that the state of the bulk matter in the bulk dual to the canonically purified state is precisely such that it gives rise to a shock in the bulk stress tensor which is required to support the Engelhardt-Wall geometry. We can view our result in two different ways. Firstly, let us assume that the bulk geometry dual to the canonically purified boundary CFT state must satisfy the semi-classical Einstein's equations, order by order in the state perturbation parameter $\cl$. In this case, the bulk must satisfy the junction conditions, equation \eqref{IJC}. Together with our result for the bulk shock, we conclude that the co-dimension two surface across which the gluing happens must satisfy 
\beq\label{QES2}
\frac{1}{4G_N}\theta_\pm + \frac{\delta S_{\text{bulk}}}{\delta x^\pm}=0,
\eeq
at $O(\cl)$, i.e., at first order in the state deformation. This is indeed the quantum extremal surface formula. On the other hand, we could assume that the gluing surface in the bulk must satisfy the quantum extremal surface formula \eqref{QES2}, without assuming that the bulk geometry satisfies the gravitational junction conditions. In this case, combining our result for the bulk stress tensor shock together with the QES formula, we would deduce the co-dimension-two junction conditions in general relativity, equation \eqref{IJC}, at first order in perturbation theory. From this point of view, the bulk gravitational equations (in this case, the junction conditions) are a consequence of the boundary entanglement structure satisfying the quantum extremal surface formula. This is in the same spirit as the results in \cite{Faulkner:2013ana, Faulkner:2017tkh, Lewkowycz:2018sgn}, but generalized now to a context where quantum corrections in the bulk are important. The quantum extremal surface formula is deeply tied-in with the structure of the bulk-to-boundary map being a quantum error correcting code, and so one might hope that this viewpoint sheds some light on the emergence of gravity from quantum error correction. It would be nice to generalize our results beyond first order in perturbation theory. One approach to do this could be to work to leading order in perturbation theory around a more general background state/geometry. We expect that with some mild assumptions on the nature of modular flow, such as approximate locality in a neighbourhood of the entanglement cut, we should be able to extend our result to this more general scenario.  

Secondly, the existence of the shock in the bulk stress tensor is deeply tied with the emergence of bulk spacetime and a correspondent quantum field theory subregion algebra in the bulk. Indeed, the calculation we presented is consistent with the expectation that the bulk state dual to the boundary canonical purification is the bulk canonical purification. From this point of view, the bulk canonical purification destroys the delicate entanglement structure at the bifurcation surface, resulting in a ``firewall''. This is in line with the results in \cite{Ceyhan:2018zfg}, where it was shown that one-sided purifications in quantum field theory can result in such shocks. In more formal terms, this is associated with the emergence of an effective type III von Neumann algebra in the bulk from the type I algebra of the boundary CFT in the large N limit \cite{Leutheusser:2021frk, Witten:2021unn, Chandrasekaran:2022eqq}. It has been recently argued in \cite{Witten:2021unn, Chandrasekaran:2022eqq} that including $1/N$ corrections, and in particular, incorporating one quantum gravitational mode (corresponding to relative time fluctuations between the two boundaries, or equivalently, one-sided mass fluctuations) changes the nature of the bulk algebra from type III to type II$_{\infty}$, thus explaining the ``renormalization'' of the UV divergence in the generalized entropy in gravity. It would be nice to understand, in a similar vein, what effect these $\frac{1}{N}$ corrections can have on the shock that we encountered, and what this means for the bulk spacetime. To this end, it would be satisfying to derive the shock from the more formal machinery of Tomita-Takesaki theory (see \cite{Witten:2018zxz} for a review). The techniques in \cite{Ceyhan:2018zfg} may be of direct relevance. Finally, it would be nice to develop more tools to study the reflection operator introduced in this paper. This would have direct applications in several useful directions in AdS/CFT such as bulk reconstruction, complexity of the bulk-to-boundary map etc.

\section*{Acknowledgements}
We thank Abhijit Gadde, Arjun Kar, Gautam Mandal, Shiraz Minwalla, Pratik Rath, Arvin Shahbazi-Moghaddam, Joan Simon, Jonathan Sorce, Sandip Trivedi and Mark Van Raamsdonk for helpful discussions and comments on the draft. We acknowledge supported from the Department of Atomic Energy, Government of India, under project identification number RTI 4002.

%\bibliographystyle{JHEP}
%\bibliography{refs}

\begin{thebibliography}{10}

\bibitem{Witten:2018zxz}
E.~Witten, \emph{{APS Medal for Exceptional Achievement in Research: Invited
  article on entanglement properties of quantum field theory}},
  \href{http://dx.doi.org/10.1103/RevModPhys.90.045003}{\emph{Rev. Mod. Phys.}
  {\bf 90} (2018) 045003}, [\href{https://arxiv.org/abs/1803.04993}{{\tt
  1803.04993}}].

\bibitem{Dutta:2019gen}
S.~Dutta and T.~Faulkner, \emph{{A canonical purification for the entanglement
  wedge cross-section}},
  \href{http://dx.doi.org/10.1007/JHEP03(2021)178}{\emph{JHEP} {\bf 03} (2021)
  178}, [\href{https://arxiv.org/abs/1905.00577}{{\tt 1905.00577}}].

\bibitem{Papadodimas:2013jku}
K.~Papadodimas and S.~Raju, \emph{{State-Dependent Bulk-Boundary Maps and Black
  Hole Complementarity}},
  \href{http://dx.doi.org/10.1103/PhysRevD.89.086010}{\emph{Phys. Rev. D} {\bf
  89} (2014) 086010}, [\href{https://arxiv.org/abs/1310.6335}{{\tt
  1310.6335}}].

\bibitem{Almheiri:2014lwa}
A.~Almheiri, X.~Dong and D.~Harlow, \emph{{Bulk Locality and Quantum Error
  Correction in AdS/CFT}},
  \href{http://dx.doi.org/10.1007/JHEP04(2015)163}{\emph{JHEP} {\bf 04} (2015)
  163}, [\href{https://arxiv.org/abs/1411.7041}{{\tt 1411.7041}}].

\bibitem{Dong:2016eik}
X.~Dong, D.~Harlow and A.~C. Wall, \emph{{Reconstruction of Bulk Operators
  within the Entanglement Wedge in Gauge-Gravity Duality}},
  \href{http://dx.doi.org/10.1103/PhysRevLett.117.021601}{\emph{Phys. Rev.
  Lett.} {\bf 117} (2016) 021601},
  [\href{https://arxiv.org/abs/1601.05416}{{\tt 1601.05416}}].

\bibitem{Harlow:2016vwg}
D.~Harlow, \emph{{The Ryu\textendash{}Takayanagi Formula from Quantum Error
  Correction}},
  \href{http://dx.doi.org/10.1007/s00220-017-2904-z}{\emph{Commun. Math. Phys.}
  {\bf 354} (2017) 865--912}, [\href{https://arxiv.org/abs/1607.03901}{{\tt
  1607.03901}}].

\bibitem{Ryu:2006bv}
S.~Ryu and T.~Takayanagi, \emph{{Holographic derivation of entanglement entropy
  from AdS/CFT}},
  \href{http://dx.doi.org/10.1103/PhysRevLett.96.181602}{\emph{Phys. Rev.
  Lett.} {\bf 96} (2006) 181602},
  [\href{https://arxiv.org/abs/hep-th/0603001}{{\tt hep-th/0603001}}].

\bibitem{Hubeny:2007xt}
V.~E. Hubeny, M.~Rangamani and T.~Takayanagi, \emph{{A Covariant holographic
  entanglement entropy proposal}},
  \href{http://dx.doi.org/10.1088/1126-6708/2007/07/062}{\emph{JHEP} {\bf 07}
  (2007) 062}, [\href{https://arxiv.org/abs/0705.0016}{{\tt 0705.0016}}].

\bibitem{Faulkner:2013ana}
T.~Faulkner, A.~Lewkowycz and J.~Maldacena, \emph{{Quantum corrections to
  holographic entanglement entropy}},
  \href{http://dx.doi.org/10.1007/JHEP11(2013)074}{\emph{JHEP} {\bf 11} (2013)
  074}, [\href{https://arxiv.org/abs/1307.2892}{{\tt 1307.2892}}].

\bibitem{Engelhardt:2018kcs}
N.~Engelhardt and A.~C. Wall, \emph{{Coarse Graining Holographic Black Holes}},
  \href{http://dx.doi.org/10.1007/JHEP05(2019)160}{\emph{JHEP} {\bf 05} (2019)
  160}, [\href{https://arxiv.org/abs/1806.01281}{{\tt 1806.01281}}].

\bibitem{Engelhardt:2021mue}
N.~Engelhardt, G.~Penington and A.~Shahbazi-Moghaddam, \emph{{A World without
  Pythons would be so Simple}},  \href{https://arxiv.org/abs/2102.07774}{{\tt
  2102.07774}}.

\bibitem{Akers:2019gcv}
C.~Akers and P.~Rath, \emph{{Entanglement Wedge Cross Sections Require
  Tripartite Entanglement}},
  \href{http://dx.doi.org/10.1007/JHEP04(2020)208}{\emph{JHEP} {\bf 04} (2020)
  208}, [\href{https://arxiv.org/abs/1911.07852}{{\tt 1911.07852}}].

\bibitem{Chandrasekaran:2020qtn}
V.~Chandrasekaran, M.~Miyaji and P.~Rath, \emph{{Including contributions from
  entanglement islands to the reflected entropy}},
  \href{http://dx.doi.org/10.1103/PhysRevD.102.086009}{\emph{Phys. Rev. D} {\bf
  102} (2020) 086009}, [\href{https://arxiv.org/abs/2006.10754}{{\tt
  2006.10754}}].

\bibitem{Zou:2020bly}
Y.~Zou, K.~Siva, T.~Soejima, R.~S.~K. Mong and M.~P. Zaletel, \emph{{Universal
  tripartite entanglement in one-dimensional many-body systems}},
  \href{http://dx.doi.org/10.1103/PhysRevLett.126.120501}{\emph{Phys. Rev.
  Lett.} {\bf 126} (2021) 120501},
  [\href{https://arxiv.org/abs/2011.11864}{{\tt 2011.11864}}].

\bibitem{Hayden:2021gno}
P.~Hayden, O.~Parrikar and J.~Sorce, \emph{{The Markov gap for geometric
  reflected entropy}},
  \href{http://dx.doi.org/10.1007/JHEP10(2021)047}{\emph{JHEP} {\bf 10} (2021)
  047}, [\href{https://arxiv.org/abs/2107.00009}{{\tt 2107.00009}}].

\bibitem{Akers:2021pvd}
C.~Akers, T.~Faulkner, S.~Lin and P.~Rath, \emph{{Reflected entropy in random
  tensor networks}},
  \href{http://dx.doi.org/10.1007/JHEP05(2022)162}{\emph{JHEP} {\bf 05} (2022)
  162}, [\href{https://arxiv.org/abs/2112.09122}{{\tt 2112.09122}}].

\bibitem{Akers:2022max}
C.~Akers, T.~Faulkner, S.~Lin and P.~Rath, \emph{{The Page curve for reflected
  entropy}}, \href{http://dx.doi.org/10.1007/JHEP06(2022)089}{\emph{JHEP} {\bf
  06} (2022) 089}, [\href{https://arxiv.org/abs/2201.11730}{{\tt 2201.11730}}].

\bibitem{Engelhardt:2022qts}
N.~Engelhardt and A.~Folkestad, \emph{{Canonical purification of evaporating
  black holes}},
  \href{http://dx.doi.org/10.1103/PhysRevD.105.086010}{\emph{Phys. Rev. D} {\bf
  105} (2022) 086010}, [\href{https://arxiv.org/abs/2201.08395}{{\tt
  2201.08395}}].

\bibitem{Geng:2020fxl}
H.~Geng, A.~Karch, C.~Perez-Pardavila, S.~Raju, L.~Randall, M.~Riojas et~al.,
  \emph{{Information Transfer with a Gravitating Bath}},
  \href{http://dx.doi.org/10.21468/SciPostPhys.10.5.103}{\emph{SciPost Phys.}
  {\bf 10} (2021) 103}, [\href{https://arxiv.org/abs/2012.04671}{{\tt
  2012.04671}}].

\bibitem{Anderson:2020vwi}
L.~Anderson, O.~Parrikar and R.~M. Soni, \emph{{Islands with gravitating baths:
  towards ER = EPR}},
  \href{http://dx.doi.org/10.1007/JHEP10(2021)226}{\emph{JHEP} {\bf 21} (2020)
  226}, [\href{https://arxiv.org/abs/2103.14746}{{\tt 2103.14746}}].

\bibitem{Marolf:2019zoo}
D.~Marolf, \emph{{CFT sewing as the dual of AdS cut-and-paste}},
  \href{http://dx.doi.org/10.1007/JHEP02(2020)152}{\emph{JHEP} {\bf 02} (2020)
  152}, [\href{https://arxiv.org/abs/1909.09330}{{\tt 1909.09330}}].

\bibitem{Engelhardt:2014gca}
N.~Engelhardt and A.~C. Wall, \emph{{Quantum Extremal Surfaces: Holographic
  Entanglement Entropy beyond the Classical Regime}},
  \href{http://dx.doi.org/10.1007/JHEP01(2015)073}{\emph{JHEP} {\bf 01} (2015)
  073}, [\href{https://arxiv.org/abs/1408.3203}{{\tt 1408.3203}}].

\bibitem{Bousso:2019dxk}
R.~Bousso, V.~Chandrasekaran and A.~Shahbazi-Moghaddam, \emph{{From black hole
  entropy to energy-minimizing states in QFT}},
  \href{http://dx.doi.org/10.1103/PhysRevD.101.046001}{\emph{Phys. Rev. D} {\bf
  101} (2020) 046001}, [\href{https://arxiv.org/abs/1906.05299}{{\tt
  1906.05299}}].

\bibitem{Lashkari:2013koa}
N.~Lashkari, M.~B. McDermott and M.~Van~Raamsdonk, \emph{{Gravitational
  dynamics from entanglement 'thermodynamics'}},
  \href{http://dx.doi.org/10.1007/JHEP04(2014)195}{\emph{JHEP} {\bf 04} (2014)
  195}, [\href{https://arxiv.org/abs/1308.3716}{{\tt 1308.3716}}].

\bibitem{Faulkner:2013ica}
T.~Faulkner, M.~Guica, T.~Hartman, R.~C. Myers and M.~Van~Raamsdonk,
  \emph{{Gravitation from Entanglement in Holographic CFTs}},
  \href{http://dx.doi.org/10.1007/JHEP03(2014)051}{\emph{JHEP} {\bf 03} (2014)
  051}, [\href{https://arxiv.org/abs/1312.7856}{{\tt 1312.7856}}].

\bibitem{Jacobson:2015hqa}
T.~Jacobson, \emph{{Entanglement Equilibrium and the Einstein Equation}},
  \href{http://dx.doi.org/10.1103/PhysRevLett.116.201101}{\emph{Phys. Rev.
  Lett.} {\bf 116} (2016) 201101},
  [\href{https://arxiv.org/abs/1505.04753}{{\tt 1505.04753}}].

\bibitem{Faulkner:2017tkh}
T.~Faulkner, F.~M. Haehl, E.~Hijano, O.~Parrikar, C.~Rabideau and
  M.~Van~Raamsdonk, \emph{{Nonlinear Gravity from Entanglement in Conformal
  Field Theories}},
  \href{http://dx.doi.org/10.1007/JHEP08(2017)057}{\emph{JHEP} {\bf 08} (2017)
  057}, [\href{https://arxiv.org/abs/1705.03026}{{\tt 1705.03026}}].

\bibitem{Haehl:2017sot}
F.~M. Haehl, E.~Hijano, O.~Parrikar and C.~Rabideau, \emph{{Higher Curvature
  Gravity from Entanglement in Conformal Field Theories}},
  \href{http://dx.doi.org/10.1103/PhysRevLett.120.201602}{\emph{Phys. Rev.
  Lett.} {\bf 120} (2018) 201602},
  [\href{https://arxiv.org/abs/1712.06620}{{\tt 1712.06620}}].

\bibitem{Lewkowycz:2018sgn}
A.~Lewkowycz and O.~Parrikar, \emph{{The holographic shape of entanglement and
  Einstein\textquoteright{}s equations}},
  \href{http://dx.doi.org/10.1007/JHEP05(2018)147}{\emph{JHEP} {\bf 05} (2018)
  147}, [\href{https://arxiv.org/abs/1802.10103}{{\tt 1802.10103}}].

\bibitem{Haehl:2019fjz}
F.~M. Haehl, E.~Mintun, J.~Pollack, A.~J. Speranza and M.~Van~Raamsdonk,
  \emph{{Nonlocal multi-trace sources and bulk entanglement in holographic
  conformal field theories}},
  \href{http://dx.doi.org/10.1007/JHEP06(2019)005}{\emph{JHEP} {\bf 06} (2019)
  005}, [\href{https://arxiv.org/abs/1904.01584}{{\tt 1904.01584}}].

\bibitem{Belin:2021htw}
A.~Belin and S.~Colin-Ellerin, \emph{{Bootstrapping quantum extremal surfaces.
  Part I. The area operator}},
  \href{http://dx.doi.org/10.1007/JHEP11(2021)021}{\emph{JHEP} {\bf 11} (2021)
  021}, [\href{https://arxiv.org/abs/2107.07516}{{\tt 2107.07516}}].

\bibitem{Israel:1966rt}
W.~Israel, \emph{{Singular hypersurfaces and thin shells in general
  relativity}}, \href{http://dx.doi.org/10.1007/BF02710419}{\emph{Nuovo Cim. B}
  {\bf 44S10} (1966) 1}.

\bibitem{Barrabes:1991ng}
C.~Barrabes and W.~Israel, \emph{{Thin shells in general relativity and
  cosmology: The Lightlike limit}},
  \href{http://dx.doi.org/10.1103/PhysRevD.43.1129}{\emph{Phys. Rev. D} {\bf
  43} (1991) 1129--1142}.

\bibitem{Czech:2017zfq}
B.~Czech, L.~Lamprou, S.~Mccandlish and J.~Sully, \emph{{Modular Berry
  Connection for Entangled Subregions in AdS/CFT}},
  \href{http://dx.doi.org/10.1103/PhysRevLett.120.091601}{\emph{Phys. Rev.
  Lett.} {\bf 120} (2018) 091601},
  [\href{https://arxiv.org/abs/1712.07123}{{\tt 1712.07123}}].

\bibitem{Czech:2018kvg}
B.~Czech, L.~Lamprou and L.~Susskind, \emph{{Entanglement Holonomies}},
  \href{https://arxiv.org/abs/1807.04276}{{\tt 1807.04276}}.

\bibitem{Czech:2019vih}
B.~Czech, J.~De~Boer, D.~Ge and L.~Lamprou, \emph{{A modular sewing kit for
  entanglement wedges}},
  \href{http://dx.doi.org/10.1007/JHEP11(2019)094}{\emph{JHEP} {\bf 11} (2019)
  094}, [\href{https://arxiv.org/abs/1903.04493}{{\tt 1903.04493}}].

\bibitem{Ceyhan:2018zfg}
F.~Ceyhan and T.~Faulkner, \emph{{Recovering the QNEC from the ANEC}},
  \href{http://dx.doi.org/10.1007/s00220-020-03751-y}{\emph{Commun. Math.
  Phys.} {\bf 377} (2020) 999--1045},
  [\href{https://arxiv.org/abs/1812.04683}{{\tt 1812.04683}}].

\bibitem{Lashkari:2019ixo}
N.~Lashkari, \emph{{Modular zero modes and sewing the states of QFT}},
  \href{http://dx.doi.org/10.1007/JHEP04(2021)189}{\emph{JHEP} {\bf 21} (2020)
  189}, [\href{https://arxiv.org/abs/1911.11153}{{\tt 1911.11153}}].

\bibitem{Bousso:2020yxi}
R.~Bousso, V.~Chandrasekaran, P.~Rath and A.~Shahbazi-Moghaddam, \emph{{Gravity
  dual of Connes cocycle flow}},
  \href{http://dx.doi.org/10.1103/PhysRevD.102.066008}{\emph{Phys. Rev. D} {\bf
  102} (2020) 066008}, [\href{https://arxiv.org/abs/2007.00230}{{\tt
  2007.00230}}].

\bibitem{Levine:2020upy}
A.~Levine, A.~Shahbazi-Moghaddam and R.~M. Soni, \emph{{Seeing the entanglement
  wedge}}, \href{http://dx.doi.org/10.1007/JHEP06(2021)134}{\emph{JHEP} {\bf
  06} (2021) 134}, [\href{https://arxiv.org/abs/2009.11305}{{\tt 2009.11305}}].

\bibitem{Marolf:2017kvq}
D.~Marolf, O.~Parrikar, C.~Rabideau, A.~Izadi~Rad and M.~Van~Raamsdonk,
  \emph{{From Euclidean Sources to Lorentzian Spacetimes in Holographic
  Conformal Field Theories}},
  \href{http://dx.doi.org/10.1007/JHEP06(2018)077}{\emph{JHEP} {\bf 06} (2018)
  077}, [\href{https://arxiv.org/abs/1709.10101}{{\tt 1709.10101}}].

\bibitem{Faulkner:2016mzt}
T.~Faulkner, R.~G. Leigh, O.~Parrikar and H.~Wang, \emph{{Modular Hamiltonians
  for Deformed Half-Spaces and the Averaged Null Energy Condition}},
  \href{http://dx.doi.org/10.1007/JHEP09(2016)038}{\emph{JHEP} {\bf 09} (2016)
  038}, [\href{https://arxiv.org/abs/1605.08072}{{\tt 1605.08072}}].

\bibitem{Sarosi:2017rsq}
G.~S\'arosi and T.~Ugajin, \emph{{Modular Hamiltonians of excited states, OPE
  blocks and emergent bulk fields}},
  \href{http://dx.doi.org/10.1007/JHEP01(2018)012}{\emph{JHEP} {\bf 01} (2018)
  012}, [\href{https://arxiv.org/abs/1705.01486}{{\tt 1705.01486}}].

\bibitem{Lashkari:2018tjh}
N.~Lashkari, H.~Liu and S.~Rajagopal, \emph{{Perturbation Theory for the
  Logarithm of a Positive Operator}},
  \href{https://arxiv.org/abs/1811.05619}{{\tt 1811.05619}}.

\bibitem{Balakrishnan:2020lbp}
S.~Balakrishnan and O.~Parrikar, \emph{{Modular Hamiltonians for Euclidean Path
  Integral States}},  \href{https://arxiv.org/abs/2002.00018}{{\tt
  2002.00018}}.

\bibitem{Rosenhaus:2014zza}
V.~Rosenhaus and M.~Smolkin, \emph{{Entanglement Entropy for Relevant and
  Geometric Perturbations}},
  \href{http://dx.doi.org/10.1007/JHEP02(2015)015}{\emph{JHEP} {\bf 02} (2015)
  015}, [\href{https://arxiv.org/abs/1410.6530}{{\tt 1410.6530}}].

\bibitem{Iyer:1994ys}
V.~Iyer and R.~M. Wald, \emph{{Some properties of Noether charge and a proposal
  for dynamical black hole entropy}},
  \href{http://dx.doi.org/10.1103/PhysRevD.50.846}{\emph{Phys. Rev. D} {\bf 50}
  (1994) 846--864}, [\href{https://arxiv.org/abs/gr-qc/9403028}{{\tt
  gr-qc/9403028}}].

\bibitem{Belin:2018fxe}
A.~Belin, A.~Lewkowycz and G.~S\'arosi, \emph{{The boundary dual of the bulk
  symplectic form}},
  \href{http://dx.doi.org/10.1016/j.physletb.2018.10.071}{\emph{Phys. Lett. B}
  {\bf 789} (2019) 71--75}, [\href{https://arxiv.org/abs/1806.10144}{{\tt
  1806.10144}}].

\bibitem{Kirklin:2019ror}
J.~Kirklin, \emph{{The Holographic Dual of the Entanglement Wedge Symplectic
  Form}}, \href{http://dx.doi.org/10.1007/JHEP01(2020)071}{\emph{JHEP} {\bf 01}
  (2020) 071}, [\href{https://arxiv.org/abs/1910.00457}{{\tt 1910.00457}}].

\bibitem{Casini:2017roe}
H.~Casini, E.~Teste and G.~Torroba, \emph{{Modular Hamiltonians on the null
  plane and the Markov property of the vacuum state}},
  \href{http://dx.doi.org/10.1088/1751-8121/aa7eaa}{\emph{J. Phys. A} {\bf 50}
  (2017) 364001}, [\href{https://arxiv.org/abs/1703.10656}{{\tt 1703.10656}}].

\bibitem{Bal}
S.~Balkrishnan and O.~Parrikar, \emph{{Shape dependence of relative entropy and
  the Connes cocyle}}, {\emph{unpublished} }.

\bibitem{Keski-Vakkuri:1998gmz}
E.~Keski-Vakkuri, \emph{{Bulk and boundary dynamics in BTZ black holes}},
  \href{http://dx.doi.org/10.1103/PhysRevD.59.104001}{\emph{Phys. Rev. D} {\bf
  59} (1999) 104001}, [\href{https://arxiv.org/abs/hep-th/9808037}{{\tt
  hep-th/9808037}}].

\bibitem{Maldacena:2001kr}
J.~M. Maldacena, \emph{{Eternal black holes in anti-de Sitter}},
  \href{http://dx.doi.org/10.1088/1126-6708/2003/04/021}{\emph{JHEP} {\bf 04}
  (2003) 021}, [\href{https://arxiv.org/abs/hep-th/0106112}{{\tt
  hep-th/0106112}}].

\bibitem{Leutheusser:2021frk}
S.~Leutheusser and H.~Liu, \emph{{Emergent times in holographic duality}},
  \href{https://arxiv.org/abs/2112.12156}{{\tt 2112.12156}}.

\bibitem{Witten:2021unn}
E.~Witten, \emph{{Gravity and the crossed product}},
  \href{http://dx.doi.org/10.1007/JHEP10(2022)008}{\emph{JHEP} {\bf 10} (2022)
  008}, [\href{https://arxiv.org/abs/2112.12828}{{\tt 2112.12828}}].

\bibitem{Chandrasekaran:2022eqq}
V.~Chandrasekaran, G.~Penington and E.~Witten, \emph{{Large N algebras and
  generalized entropy}},  \href{https://arxiv.org/abs/2209.10454}{{\tt
  2209.10454}}.

\end{thebibliography}
\providecommand{\href}[2]{#2}\begingroup\raggedright\endgroup

\end{document}